\documentclass[prb,aps,twocolumn,superscriptaddress,showpacs]{revtex4-1}

\makeatletter
\renewcommand*{\@fnsymbol}[1]{\ensuremath{\ifcase#1\or \dagger\or \dagger\or \ddagger\or
		\mathsection\or \mathparagraph\or \|\or **\or \dagger\dagger
		\or \ddagger\ddagger \else\@ctrerr\fi}}
\makeatother

\usepackage{graphicx}% Include figure files
\usepackage{epstopdf}
\usepackage[colorlinks=true,linkcolor=black, citecolor=blue, urlcolor=blue, 
    unicode=true]{hyperref}

\usepackage{color}
\usepackage{hyperref} 
\usepackage{anysize}
\usepackage{hhline,multirow,float}
\usepackage{makecell}
\usepackage{array}
\usepackage{dcolumn}
\newcolumntype{?}{!{\vrule width 1pt}}
\usepackage{color}

\definecolor{model1}{RGB}{255,0,0}
\definecolor{model2}{RGB}{0,176,80}
\definecolor{model3}{RGB}{0,0,255}
\definecolor{model4}{RGB}{255,165,0}

\usepackage{multirow}
\usepackage[table]{xcolor}
\usepackage{capt-of}
\usepackage{wasysym}

\marginsize{1.7 cm}{1.7 cm}{0.5 cm}{0.5 cm}

\newcommand{\QQ}{$|\mathbf{Q}|$}

\newcommand{\mgcoo}{Co$_{0.03}$Mg$_{0.97}$O}

\begin{document}
\newcommand\bbone{\ensuremath{\mathbbm{1}}}
\newcommand{\ul}{\underline}
\newcommand{\bp}{{\bf p}}
\newcommand{\vl}{v_{_L}}
\newcommand{\vc}{\mathbf}
\newcommand{\be}{\begin{equation}}
\newcommand{\ee}{\end{equation}}
\newcommand{\bk}{{{\bf{k}}}}
\newcommand{\bK}{{{\bf{K}}}}
\newcommand{\cE}{{{\cal E}}}
\newcommand{\bQ}{{{\bf{Q}}}}
\newcommand{\br}{{{\bf{r}}}}
\newcommand{\bg}{{{\bf{g}}}}
\newcommand{\bG}{{{\bf{G}}}}
\newcommand{\hbr}{{\hat{\bf{r}}}}
\newcommand{\bR}{{{\bf{R}}}}
\newcommand{\bq}{{\bf{q}}}
\newcommand{\hx}{{\hat{x}}}
\newcommand{\hy}{{\hat{y}}}
\newcommand{\hd}{{\hat{\delta}}}
\newcommand{\bea}{\begin{eqnarray}}
\newcommand{\eea}{\end{eqnarray}}
\newcommand{\ra}{\rangle}
\newcommand{\la}{\langle}
\renewcommand{\tt}{{\tilde{t}}}
\newcommand{\upa}{\uparrow}
\newcommand{\dna}{\downarrow}
\newcommand{\bS}{{\bf S}}
\newcommand{\vS}{\vec{S}}
\newcommand{\dg}{{\dagger}}
\newcommand{\pdg}{{\phantom\dagger}}
\newcommand{\tphi}{{\tilde\phi}}
\newcommand{\cf}{{\cal F}}
\newcommand{\ca}{{\cal A}}
\renewcommand{\ni}{\noindent}
\newcommand{\ct}{{\cal T}}
\newcommand{\brf}{\bar{F}}
\newcommand{\brg}{\bar{G}}
\newcommand{\jeff}{j_{\rm eff}}

\title{Disentangling orbital and spin exchange interactions for Co$^{2+}$ on a rocksalt lattice}

\author{P.~M.~Sarte}
\affiliation{School of Chemistry, University of Edinburgh, Edinburgh EH9 3FJ, United Kingdom}
\affiliation{Centre for Science at Extreme Conditions, University of Edinburgh, Edinburgh EH9 3FD, United Kingdom}
\author{R.~A.~Cowley} \altaffiliation{Deceased 27 January 2015}
\affiliation{Department of Physics, Clarendon Laboratory, University of Oxford, Park Road, Oxford, OX1 3PU, United Kingdom}
\author{E.~E.~Rodriguez} 
\affiliation{Department of Chemistry and Biochemistry, University of Maryland, College Park, Maryland 20742, USA} 
\author{E.~Pachoud} 
\affiliation{School of Chemistry, University of Edinburgh, Edinburgh EH9 3FJ, United Kingdom}
\affiliation{Centre for Science at Extreme Conditions, University of Edinburgh, Edinburgh EH9 3FD, United Kingdom}
\author{D.~Le}
\affiliation{ISIS Facility, Rutherford Appleton Laboratory, Chilton, Didcot OX11 0QX, United Kingdom}
\author{V.~Garc\'{i}a-Sakai}
\affiliation{ISIS Facility, Rutherford Appleton Laboratory, Chilton, Didcot OX11 0QX, United Kingdom}
\author{J.~W. Taylor} 
\affiliation{ISIS Facility, Rutherford Appleton Laboratory, Chilton, Didcot OX11 0QX, United Kingdom}
\author{C.~D.~Frost} 
\affiliation{ISIS Facility, Rutherford Appleton Laboratory, Chilton, Didcot OX11 0QX, United Kingdom}
\author{D.~Prabhakaran} 
\affiliation{Department of Physics, Clarendon Laboratory, University of Oxford, Park Road, Oxford, OX1 3PU, United Kingdom}
\author{C.~MacEwen} 
\affiliation{School of Physics and Astronomy, University of Edinburgh, Edinburgh EH9 3FD, United Kingdom}
\author{A.~Kitada}
\affiliation{Department of Materials Science and Engineering, Kyoto University, Yoshida-honmachi, Sakyo, Kyoto 606-8501, Japan}
\author{A.~J.~Browne} 
\affiliation{School of Chemistry, University of Edinburgh, Edinburgh EH9 3FJ, United Kingdom}
\affiliation{Centre for Science at Extreme Conditions, University of Edinburgh, Edinburgh EH9 3FD, United Kingdom}
\author{M.~Songvilay}
\affiliation{Centre for Science at Extreme Conditions, University of Edinburgh, Edinburgh EH9 3FD, United Kingdom}
\affiliation{School of Physics and Astronomy, University of Edinburgh, Edinburgh EH9 3FD, United Kingdom}
\author{Z. Yamani}
\affiliation{National Research Council, Chalk River, Ontario K0J 1JO, Canada }
\author{W.~J.~L.~Buyers}
\affiliation{National Research Council, Chalk River, Ontario K0J 1JO, Canada }
\affiliation{Canadian Institute of Advanced Research, Toronto, Ontario M5G 1Z8, Canada}
\author{J.~P.~Attfield}
\affiliation{School of Chemistry, University of Edinburgh, Edinburgh EH9 3FJ, United Kingdom}
\affiliation{Centre for Science at Extreme Conditions, University of Edinburgh, Edinburgh EH9 3FD, United Kingdom}
\author{C.~Stock}
\affiliation{Centre for Science at Extreme Conditions, University of Edinburgh, Edinburgh EH9 3FD, United Kingdom}
\affiliation{School of Physics and Astronomy, University of Edinburgh, Edinburgh EH9 3FD, United Kingdom}

\date{\today}

\begin{abstract}

Neutron spectroscopy was applied to study the magnetic interactions of orbitally degenerate Co$^{2+}$ on a host MgO rocksalt lattice where no long range spin or orbital order exists.  The paramagnetic nature of the substituted monoxide Co$_{0.03}$Mg$_{0.97}$O allows for the disentanglement of spin-exchange and spin-orbit interactions.  By considering the prevalent excitations from Co$^{2+}$ spin pairs, we extract 7 exchange constants out to the fourth coordination shell.  An antiferromagnetic next nearest neighbor 180$^{\circ}$ exchange interaction is dominant, however dual ferromagnetic and antiferromagnetic interactions are observed for pairings with other pathways.  These interactions can be understood in terms of a combination of orbital degeneracy in the $t_{2g}$ channel and the Goodenough-Kanamori-Anderson (GKA) rules. Our work suggests that such a hierarchy of exchange interactions exists in transition metal-based oxides with a $t_{2g}$ orbital degeneracy.

\end{abstract}

\maketitle

\section{Introduction:}

The combination of magnetic exchange and orbital degeneracy has provided the basis for a number of topics in condensed matter physics including metal-insulator transitions, high temperature superconductors, colossal magnetoresistance,~\cite{Tokura,Dagotto,Kugel}  and more recently Kitaev interactions~\cite{Okamoto07,Wang17,Jackeli09}.  Rocksalt CoO was the first orbitally degenerate compound to have its magnetic structure investigated using neutron diffraction,~\cite{Shull1951,Li1955,structure1,structure2} but the underlying exchange interactions are still not known.  Indeed, calculations and experiment have been hindered by the complex electronic and orbital ground state of Co$^{2+}$.  While $e_{g}$ mediated magnetic exchange has been well understood (for example in KCuF$_{3}$~\cite{Satija}), the case of exchange involving degenerate $t_{2g}$ orbitals has proven more difficult.~\cite{Oles}  We investigate the magnetic exchange interactions in the case of a $t_{2g}$ orbital degeneracy by performing neutron spectroscopy on MgO substituted with Co$^{2+}$.  We extract 7 exchange interactions and observe dual ferro and antiferromagnetic exchange interactions with comparable magnitudes.  The dual exchange interactions is a direct result of the underlying $t_{2g}$ orbital degeneracy of Co$^{2+}$.

\begin{figure}[h!]
	\centering
	\includegraphics[width=1.0\linewidth]{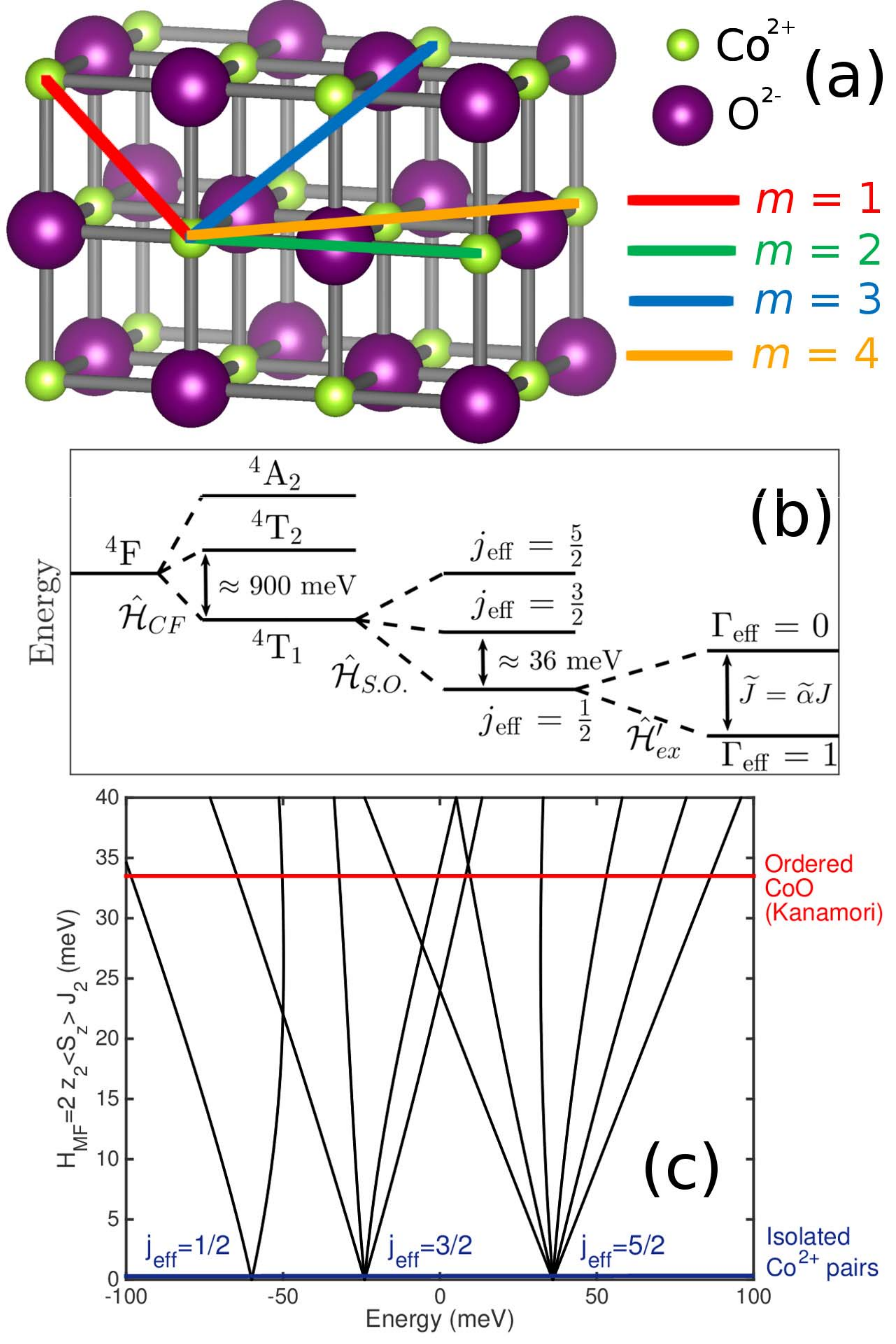}
	\caption{$(a)$ Cubic (room temperature) rock salt $Fm\overline{3}m$ crystal structure of CoO~\cite{jauch}. The pair distances between first shell (nearest) neighbors, second shell (next nearest) neighbors, \emph{etc}. are denoted by $m$ = 1, 2, \emph{etc}., respectively. $(b)$ the effective pair Hamiltonian $\hat{\mathcal{H}}_{pair}$ for \mgcoo. $(c)$ The energy eigenvalues of the single ion Hamiltonian including a molecular field from magnetic order with Kanamori's estimate~\cite{1957} of $J_{2}$ shown by the solid red line.}
	\label{fig:fig1}
\end{figure}

The starting point for understanding the spin-orbital Hamiltonian for paramagnetic Co$^{2+}$ ions is crystal field theory based on octahedral coordination~\cite{haverkort2007,larson2007} (Fig.~\ref{fig:fig1}$(a)$ for rocksalt CoO)~\cite{sakurai,1957,cowley}. As schematically shown in Fig.~\ref{fig:fig1}$(b)$, the octahedral crystal field ($\hat{\mathcal{H}}_{CF}$) splits the 5 degenerate $d$-orbitals ($^{4}F$) such that the  $d^{7}$ electronic structure consists of 5 electrons occupying the lower energy $t_{2g}$ states and 2 electrons in the higher energy $e_{g}$ orbitals.  This constitutes two orbital triplets ($^4T_{1,2}$) levels separated by $10Dq$~$\sim$~900~meV~\cite{cowley}.  The triplet degenerate $^4T_{1}$ ground state can be approximated to have an effective orbital angular momentum of $\widetilde{l}$~=~1.~\cite{cowley,cov2o6,projection,projection2,1957,1957_2,buyers,buyers1984,cowley2} Applying spin-orbit coupling (defined by $\hat{\mathcal{H}}_{S.O.}=\tilde{\lambda} \vec{l}\cdot\vec{S}$, with $S=\frac{3}{2}$) to this orbital ground state results in 3 effective spin-orbit manifolds classified by an effective angular momentum of $j\rm_{{eff}}$~=~$\frac{1}{2}$, $\frac{3}{2}$ and $\frac{5}{2}$ (with $\vec{j}\rm{_{eff}}$$=\vec{l}+\vec{S}$).   The $j\rm_{_{eff}}=\frac{1}{2}$ ground state is separated from the higher energy $j\rm_{_{eff}}=\frac{3}{2}$ states by $\frac{3}{2}~\tilde{\lambda}\sim$~36~meV~\cite{cowley}. 

In the presence of long range magnetic order (as exists in CoO at low temperatures), the total single-ion Hamiltonian for Co$^{2+}$ can then be summarized by 

\begin{equation}
\hat{\mathcal{H}}_{SI} = \hat{\mathcal{H}}_{CF} + \hat{\mathcal{H}}_{S.O.}  + \hat{\mathcal{H}}_{MF}
\end{equation}

\noindent where $\hat{\mathcal{H}}_{CF}$, $\hat{\mathcal{H}}_{S.O.}$ and $\hat{\mathcal{H}}_{MF}$ are the octahedral crystal field, spin-orbit, and magnetic order induced molecular field. The effect of magnetic ordering on the three spin-orbit manifolds discussed above can be illustrated by considering a single dominant next nearest neighbor 180$\rm{^{o}}$ Co$^{2+}$-O$^{2-}$-Co$^{2+}$ superexchange $J_{2}$ with  

\begin{equation}
\hat{\mathcal{H}}_{MF}=2J_{2}z_{2}\langle \hat{\mathcal{\mathbf{S}}} \rangle_{av}\hat{\mathcal{S}}^{z}, 
\end{equation}

\noindent where $z_{2}$ and $\hat{\mathcal{S}}^{z}$ denote the number of Co$^{2+}$ neighbors and the $z$-axis of the spin operator~\cite{sakurai}.  As illustrated in Fig.~\ref{fig:fig1}$(c)$, by considering only the predicted value of $J_{2}$ by Kanamori~\cite{1957} in the mean field expression for $\hat{\mathcal{H}}_{MF}$, a complex admixture of different molecular field split Co$^{2+}$ spin-orbit manifolds occurs in the presence of magnetic order~\cite{1957,cowley,sakurai}.   

The strong magnetic ordered induced mixing of multiple $j\rm_{_{eff}}$ manifolds in CoO is in contrast to many other Co$^{2+}$ based magnets that have both weak exchange and molecular fields and thus exhibit weak mixing~\cite{cov2o6,coldea,grenier,zhou2012}.  CoO is further complicated by the possibility of multiple long-range spin-spin interactions~\cite{Dalverny,2011,tomiyasu,DFT}.  The extraction of the multiple spin exchanges in CoO is thus experimentally very difficult despite the simplicity of its crystal structure~\cite{buyers,buyers1984,2011,cowley,sakurai,yamani2,fischer2,tomiyasu,kant}.  

We have extracted the magnetic exchange interactions on a rocksalt lattice by investigating weakly substituted \mgcoo~using neutron scattering and through considering excitations from the dominant Co$^{2+}$ pair response.  This paper is divided into four sections including this introduction.  We first describe the experimental methods including materials preparation and characterization techniques where we conclude that our dilute sample can be described by a Co$^{2+}$ pair response.  An expanded description of the characterization is given in the Supplementary Information illustrating the x-ray, susceptibility, and EDX data.~\cite{suppl}  In section three, the theory required to extract both the exchange constant and also the distance associated with the interaction is outlined.  We then show the experimental data used to derive the exchange interactions.  We finally conclude with a discussion of the results including a comparison with thermodynamic data from pure CoO and also how we can understand the results in terms of the GKA rules.

\section{Experimental details and materials characterization:}

To extract individual $J$ constants for Co$^{2+}$, we have followed the pioneering work on dilute Mn$^{2+}$~\cite{furrer2,Svensson} and Co$^{2+}$~\cite{buyers} compounds and measured the dilute monoxide \mgcoo~using inelastic neutron spectroscopy.    The high magnetic dilution removes the problematic molecular field discussed above (Fig.~\ref{fig:fig1}$(c)$) and suppresses the mixing between $j_{eff}$ manifolds allowing us to consider a dominant response for Co$^{2+}$ pairs.  Probabilistic arguments can be used to illustrate this and is based on the observation that for a given random distribution of $x$ Co$^{2+}$ and $(1-x)$ Mg$^{2+}$ ions, the number of Co$^{2+}$ pairs and the number of pairwise interactions for a given geometry present in the lattice far outweighs the number of Co$^{2+}$ triplets and corresponding interactions between three Co$^{2+}$ cations. For example, if there are $N$ ways that a cluster with a particular geometry of 3 sites $XYZ$ can occur in a given crystal, the relative probabilities of an arrangement of 3 Mg$^{2+}$, 1 Co$^{2+}$ and 2 Mg$^{2+}$ (and its permutations), 2 Co$^{2+}$ and 1 Mg$^{2+}$ (and its permutations) and 3 Co$^{2+}$ occupying the three sites $XYZ$ are $(1-x)^3$, $x(1-x)^2$, $x^2(1-x)$ and $x^3$, respectively. Hence the ratio of numbers of spin pairs with $XY$, $XZ$ and $YZ$ geometry to spin triplets with $XYZ$ geometry in the lattice is $=~\frac{1-x}{x}$, and thus for small $x$, the number and hence inelastic neutron scattering intensities of Co$^{2+}$ pair excitations far outweigh those from larger Co$^{2+}$ clusters.  We summarize the sample preparation and characterization techniques confirming the dominant pair response in this section and expanded description, including data from the techniques, of the characterization is provided in the Supplementary Information.  We also discuss the neutron experiments applied to these materials.

\textit{Materials Preparation:} Two polycrystalline samples of \mgcoo~were synthesized for this particular investigation. The first was synthesized by traditional solid state methods as outlined by Cowley \emph{et al.}~\cite{cowley}. A second sample of \mgcoo~was made using solution techniques by mixing stoichiometric amounts of Mg(NO$_{3}$)$_{2}$~$\cdot$~6H$_{2}$O and Co(NO$_{3}$)$_{2}$~$\cdot$~6H$_{2}$O. The solid mixture was dissolved in CH$_{3}$CH$_{2}$OH and stirred for 1 h and heated to 70$\rm{^{o}}$C for 12~h yielding a pink gel.  The gel was heated in air to 600$\rm{^{o}}$C with a heating rate of 20$\rm{^{o}}$C/h, reacted for 24~h, subsequently heated to 1000$\rm{^{o}}$C with a heating rate of 150$\rm{^{o}}$C/h, held for an additional 48~h and finally cooled to room temperature by switching off the furnace.. Details concerning the synthesis and treatment of MgO and CoO samples are outlined by Cowley \emph{et al.}~\cite{cowley}.  We note that both magnetically substituted MgO samples gave consistent results and the comparison is shown in the Supplementary Information.

\indent \textit{Laboratory X-ray Diffraction:} \indent Room temperature powder diffraction patterns of the end members (CoO and MgO) and Co$_{x}$Mg$_{1-x}$O synthesized by sol-gel were collected over 2$\theta$ = [25,100]$^{\circ}$ in 0.02$^{\circ}$ steps on a Bruker D2 Phaser laboratory x-ray diffractometer utilizing a monochromated Cu K$_{\alpha,1,2}$ source. As illustrated in the Supplementary Information, Rietveld refinement of Mg$_{1-x}$Co$_{x}$O indicates that the solid solution assumes a rocksalt structure ($Fm\bar{3}m$) with a unit cell parameter $a$~=~4.2131(2) \AA. Utilizing the measured values of the end members: CoO (4.2594(4)~\AA) and MgO (4.2118(1)~\AA), the unit cell parameter of 4.2131(2) \AA~corresponds to an $x$~=~0.025(5) according to Vegard's law~\cite{vegard}, supporting that approximately 3\% of the Mg$^{2+}$ sites contain Co$^{2+}$. 

\indent \textit{Energy dispersive x-ray analysis:} As a final direct confirmation of the concentration of Co$^{2+}$ in our sample we performed energy dispersive x-ray measurements.  Elemental analysis was performed using scanning electron microscopy (SEM) on a Hitachi SU-70 Schottky field emission gun SEM with an equipped Bruker Quantax energy dispersive X-ray detector. Energy dispersive X-ray spectroscopy (EDS) was carried out at 15 keV.  The results are illustrated in Supplementary Information show the effect substitution and the homogeneous distribution of cobalt throughout the sample.

\indent \textit{DC Magnetic Susceptibility:}  Temperature dependence of magnetization was measured on a Quantum Design MPMS for a 32.5 mg of polycrystalline \mgcoo~synthesized by sol-gel in an external DC field $\rm{\mu_{o}H_{ext}}$~=~0.1~T. ZFC measurements were performed in 2~K steps spaced linearly from 2~K to 300~K, while FC measurements were performed in 5~K steps spaced linearly from 2~K to 170~K. For both ZFC/FC measurements.   As described in the Supplementary Information, the Curie-Weiss constant was found to be consistent with pairs of Co$^{2+}$ with an exchange interaction reported by Kanamori.~\cite{1957}  The Curie constant was found to agree with a concentration of Co$^{2+}$ ions consistent with starting concentrations, x-ray powder diffraction, and also EDX measurements.  Susceptibility measurements therefore confirm the following key experimental properties of our substituted samples: the lack of magnetic ordering; the absence of measurable clustering of Co$^{2+}$ evidenced from no measurable difference between zero-field and field-cooled sweeps; a Curie-Weiss constant consistent with a dominant 180$^{\circ}$ superexchange interaction; and finally a Curie constant consistent with starting concentrations.

%Based on x-ray diffraction, EDX, and susceptibility we conclude that our samples of MgO weakly substituted with Co$^{2+}$ ions can be considered in terms of isolated Co$^{2}$ pairs.  

\indent \textit{Inelastic Neutron Scattering Details:} 45.8~g, 45.2~g, 32.5~g and 15.7~g of \mgcoo~synthesized by the standard solid state and sol-gel methods, annealed MgO and CoO, respectively, were placed in separate airtight aluminum cans under helium.  The high-energy measurements were made on the direct geometry MARI spectrometer. For measurements concerning the \mgcoo~sample synthesized by traditional solid state methods, MgO and CoO powders, the $t_{o}$ chopper was operated at 50 Hz  in parallel with a Gd chopper spun at frequencies $f$ = 350, 250 and 150 Hz with incident energies E$\rm{_{i}}$~=~30,~10~and~5~meV, respectively, providing an elastic resolution of 0.7, 0.2 and 0.1~meV, respectively. For measurements concerning the \mgcoo~sample synthesized by sol-gel, the Gd chopper was spun at $f$~=~350~Hz and 250~Hz with an E$\rm{_{i}}$ of 29.50~meV and 14.50~meV, providing an elastic resolution of 0.7~meV and 0.2~meV, respectively. For both \mgcoo~samples, a thick disk chopper with $f$~=~50~Hz reduced the background from high-energy neutrons. A top loading Displex CCR cooled the samples to a base temperature of approximately 5~K.   We note that further neutron inelastic scattering results comparing pure MgO, CoO, and our substituted MgO sample are presented in the supplementary information.

\indent For lower energies, measurements were made on the indirect geometry IRIS spectrometer. The final energy was fixed at 1.84~meV by PG002 analyzer crystals in near backscattering geometry. The graphite analyzers are cooled to reduce thermal diffuse scattering, providing an elastic resolution of 17.5~$\mu$eV. A combination of IRIS' long path length and its array of disc choppers, allowed us to select multiple time windows, resulting in the measured bandwidth being selectively increased to include energy transfers up to $\sim$ 2 meV. A top loading displex CCR was used to cool the sample to a base temperature of approximately 11 K. For all samples, identical instrumental and environmental parameters were employed on IRIS.

\begin{figure}
	\centering
	\includegraphics[width=1.0\linewidth]{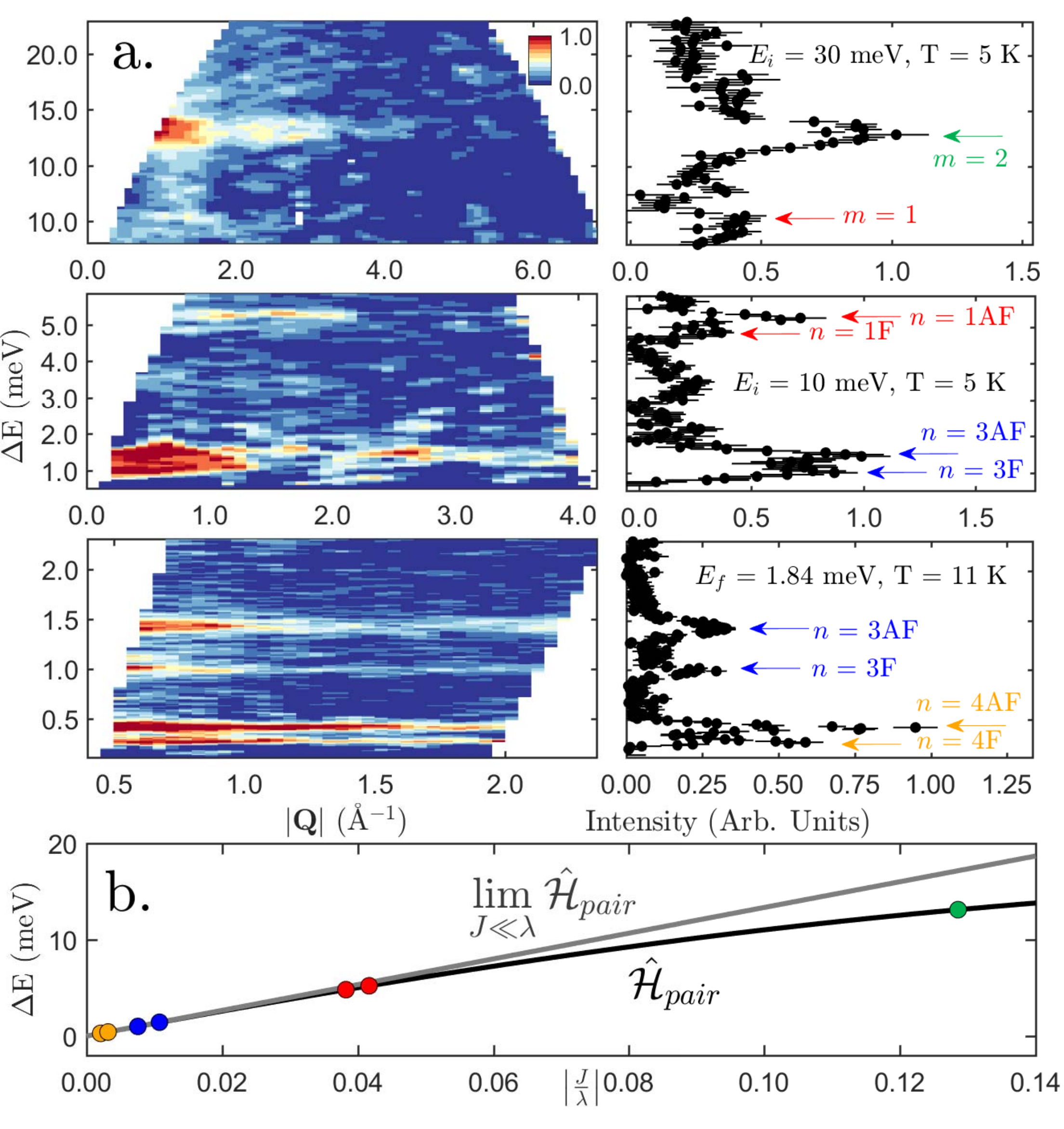}
	\caption{ $(a)$ Background (using pure and non magnetic MgO) subtracted powder averaged neutron scattering intensity maps of  \mgcoo~measured on ($a$, top left) MARI at 5~K with an E$\rm{_{i}}$~=~30~meV, ($a$, middle left) MARI at 5~K with an E$_{i}$ = 10~meV and ($a$, bottom left) IRIS at 11~K with an E$\rm_{_{f}}$ of 1.84~meV revealing 7 low energy bands of dispersionless magnetic excitations.  The right column shows $|\mathbf{Q}|$-integrated cuts. Labels denote the coordination shell $m$ and the type of coupling present with label $n$, both of which are determined in Fig.~\ref{fig:fig3}. $(b)$ The black curve denotes the pair energy splitting as a function of the normalized exchange $\Delta E\left(\left|\frac{J}{\lambda}\right|\right)$. The points are measured energy positions from $(a)$.  The grey line is the same relationship derived using the projection theorem in the large $\lambda$ limit~\cite{projection,projection2}.}
	\label{fig:fig2}
\end{figure}

\begin{figure*}
	\centering
	\includegraphics[width=1.0\linewidth]{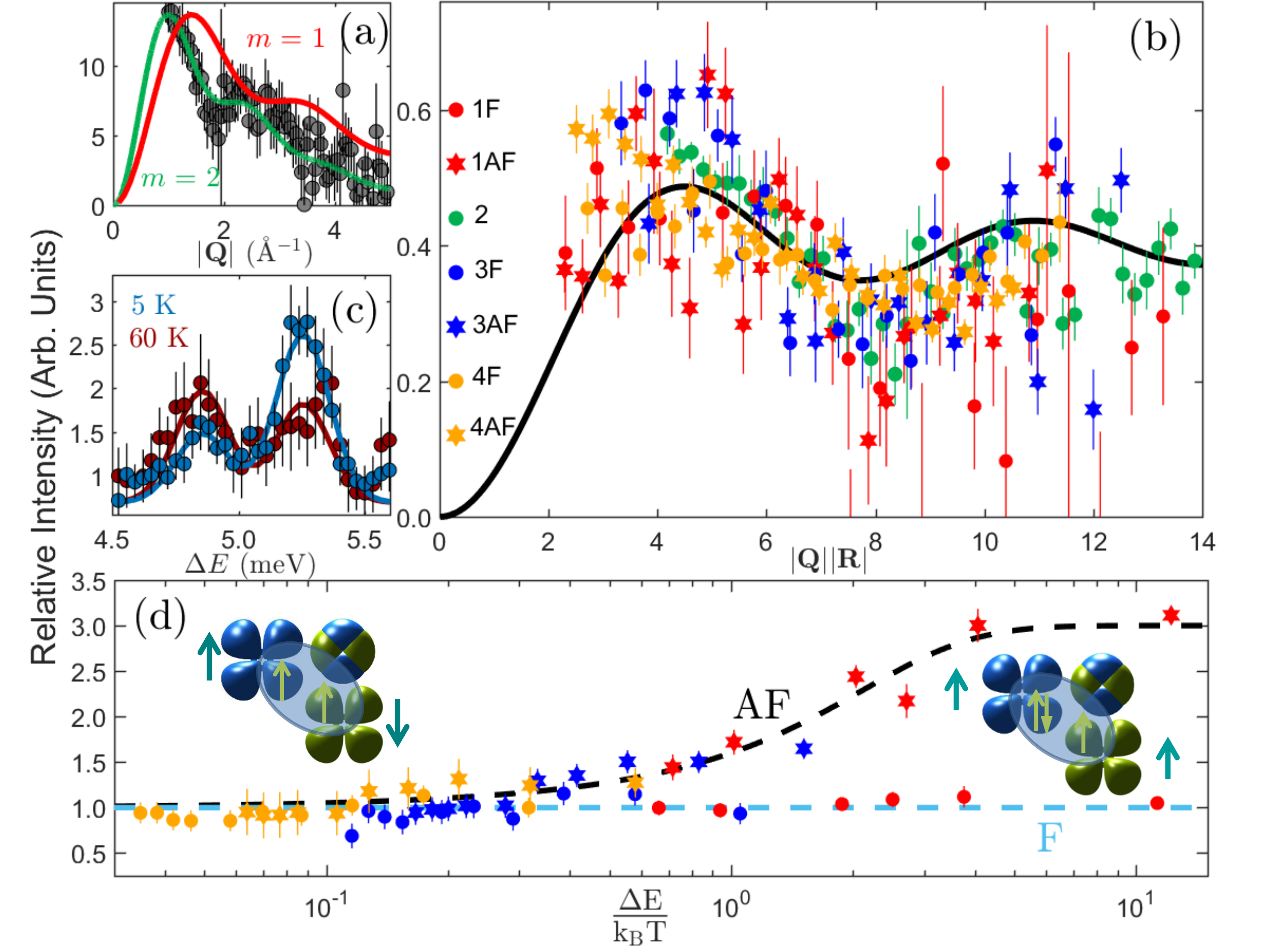}
	\caption{$(a)$ Constant-$E$ cut ($\Delta E$ = [12,14] meV) from MARI at 5 K with an E$\rm_{_{i}}$=30 meV. The green curve is a fit to Eq.~\ref{eq:4} with $|\mathbf{R}|$~=~4.2(3)~\AA\ ($m$~=~2 pairs). The red curve is with $|\mathbf{R}|$ fixed as 2.98~\AA\ ($m$~=~1 pairs). $(b)$ Scaled and form factor corrected \QQ-dependence of the intensities for all magnetic excitations with $|\mathbf{R}|$ calculated from the fitting routine described in $(a)$.  The solid black curve is $1-\frac{\sin(|\mathbf{Q}||\mathbf{R}_{n}|)}{|\mathbf{Q}||\mathbf{R}_{n}|}$. $(c)$ Constant-$|\mathbf{Q}|$ cut (MARI, $E\rm_{_{i}}$=10 meV) showing a different temperature dependence for the two peaks despite both being from $m$~=~1 pairs. $(d)$ Normalized temperature dependence of the Bose-factor corrected integrated intensity for all 7 excitations (Fig.~\ref{fig:fig2}) showing two universal curves calculated (dashed lines) for antiferromagnetic and ferromagnetic coupling. Both the integrated intensities and the calculated behaviour of antiferromagnetic or ferromagnetically coupled pairs were normalized by $I_{F}(T)$ as described in the main text. The inset is a pictorial representation of the sign of $J$ as predicted by the GKA rules~\cite{rules2,rules3,rules4} --- antiferromagnetism (\textbf{left}) is a result of exchange between two half-filled t$_{2g}$ orbitals while weaker ferromagnetism (\textbf{right}) is a result of exchange between a half-filled and completely filled t$_{2g}$ orbitals. Yellow arrows denote local $t_{2g}$ spin configurations and teal arrows denote total spin configurations on each Co$^{2+}$.  }
	\label{fig:fig3}
\end{figure*}

\section{Co$^{2+}$ pair interactions:}

\indent Having discussed the materials preparation and characterization, we conclude that our rocksalt MgO sample substituted with Co$^{2+}$ can be considered to be dominated by pairs of Co$^{2+}$ ions.  We now discuss the neutron scattering response of an isolated pair of magnetic ions and how it can be used to extract both the interaction distance and also the energy exchange interaction.  By considering Co$^{2+}$ pair interactions and only low energy excitations within the lowest $j\rm_{_{eff}}=\frac{1}{2}$ doublet (with $\hat{\tilde{\mathbf{j}}}=\beta \hat{\mathbf{S}}$), the interaction energy $\hat{\mathcal{H}}_{ex}$ between a pair of Co$^{2+}$ ions in substituted Mg$_{0.97}$Co$_{0.03}$O, is approximated by

\begin{equation}
\hat{\mathcal{H}}'_{ex} = 2J \hat{\bf{S}}_{1}\cdot \hat{\bf{S}}_{2} \sim \widetilde{\alpha} J\  \hat{\widetilde{\mathbf{j}}}_{1} \cdot \hat{\widetilde{\mathbf{j}}}_{2},
\label{eq:3}
\end{equation}

\noindent where $\hat{\widetilde{\mathbf{j}}}$ and $\widetilde{\alpha}=2\beta^{2}$ denotes an effective total angular momentum operator with $j=\frac{1}{2}$ and a projection factor, respectively. As summarized by Fig.~\ref{fig:fig1}$(b)$, the $\hat{\mathcal{H}}'_{ex}$ describes individual $j\rm{_{eff}} = \frac{1}{2}$ pair excitations as transitions between triplet ($\Gamma\rm_{{eff}}$~=~1) and singlet ($\Gamma\rm_{_{eff}}$~=~0) levels separated by an energy of $\Delta E$~=~$\widetilde{\alpha}J$~\cite{dimer,projection3,furrer}.  The projection factor $\widetilde{\alpha}$, in this low energy approximation, can be calculated by diagonalizing $\hat{\mathcal{H_{S.I.}}}+\hat{\mathcal{H}}'_{ex}$ with $\hat{\mathcal{H_{MF}}}~=~0$ owing to the lack of long range magnetic order in \mgcoo~\cite{cowley}.  This is equivalent to the following Hamiltonian for two (labelled 1 and 2) interacting Co$^{2+}$ ions,

\begin{equation}
\hat{\mathcal{H'}}= \widetilde{\lambda}\hat{\widetilde{\mathbf{l}}}_{1} \cdot \hat{\mathbf{S}}_{1} + \widetilde{\lambda}\hat{\widetilde{\mathbf{l}}}_{2} \cdot \hat{\mathbf{S}}_{2} + 2J \hat{\mathcal{\mathbf{S}}}_{1} \cdot \hat{\mathcal{\mathbf{S}}}_{2}.
\end{equation}

\noindent By considering $\tilde{l}=1$ and $S=\frac{3}{2}$, this amounts to 144 basis states and a 144~$\times$~144 matrix for this particular Hamiltonian in terms of the two particle basis of $|\widetilde{l}_{1}, m_{\widetilde{l},1}, s_{1}, m_{s,1}\rangle$ $\otimes$ $|\widetilde{l}_{2}, m_{\widetilde{l},2}, s_{2}, m_{s,2}\rangle$, where $\widetilde{l}_{i}$, $m_{\widetilde{l},i}$, $s_{i}$, $m_{s,i}$ denote the eigenvalues corresponding to the $\hat{\widetilde{\mathbf{l}}}_{i}$, $\hat{\widetilde{\mathbf{l}}}_{z,i}$, $\hat{\mathbf{S}}_{i}$ and $\hat{\mathbf{S}}_{z,i}$ operators, respectively, for the $i$\textsuperscript{th} particle.  As illustrated in Fig.~\ref{fig:fig2}$(b)$, in the limit of $J \ll \lambda$, $\Delta E(J)$ is linear with  $\widetilde{\alpha}=\frac{50}{9}$ in agreement with the projection theorem of angular momentum~\cite{projection2,projection3}.  Therefore, measuring pair excitations with neutron spectroscopy provides a direct way to estimate the magnitude of exchange constant $|J|$ between neighboring Co$^{2+}$ ions when this projection factor is taken into account.  We note that this is independent of the sign of $J$ and we discuss how that can be determined from the temperature dependence below.

\begin{table*}
	\caption{Magnetic exchange constants for \mgcoo~determined by the current study, magnetic exchange constants for CoO as cited in literature~\cite{2011,tomiyasu} and calculated for CoO by Deng \emph{et al.}~\cite{DFT} using $GGA+U$ DFT. The values from $GGA+U$ DFT have been renormalized such that $J_{2}$ is equal to the value from this current study. The values of T$_{N}$, $\theta\rm_{_{CW}}$ and $\lambda$ reported in literature~\cite{jauch,weiss,cowley,singer} for CoO have been included for the purposes of a comparison to the mean field value~\cite{MFtheory,kittel} of $\theta\rm_{_{CW}}$ corresponding to the $J$ values determined by the current study.}
\begin{ruledtabular}
	\begin{tabular}{cccc}
		%\Xhline{3\arrayrulewidth} 
		Quantity $\biggl/$ Source & Current Study (meV)  & Literature Studies (meV)  & Calculated (meV)~\cite{DFT}   \\ 
		\Xhline{3\arrayrulewidth}
		$\widetilde{\lambda}$ &  & 24(5)~\cite{cowley} &   \\
		\Xhline{3\arrayrulewidth}
		\textcolor{model1}{$J_{1AF}$} & \textcolor{model1}{1.000(8)} & \multirow{ 2}{*}{\textcolor{model1}{0.60 to -0.31}~\cite{2011,1957}} & \multirow{ 2}{*}{\textcolor{model1}{-0.97(2)}} \\ 
		\textcolor{model1}{$J_{1F}$} & \textcolor{model1}{-0.918(6)}  &    & \\ 
		\Xhline{3\arrayrulewidth}
		\textcolor{model2}{$J_{2}$ or $J_{2AF}$} & \textcolor{model2}{3.09(5)}  & \textcolor{model2}{2.8 to 0.0013}~\cite{2011}   & \textcolor{model2}{3.09(5)} \\ \Xhline{3\arrayrulewidth}
		\textcolor{model3}{$J_{3AF}$} & \textcolor{model3}{0.258(1)} & \multirow{ 2}{*}{\textcolor{model3}{-0.67}~\cite{tomiyasu}}   & \multirow{ 2}{*}{\textcolor{model3}{-0.461(8)}} \\ 
		\textcolor{model3}{$J_{3F}$} & \textcolor{model3}{-0.182(1)} &  & \\  
		\Xhline{3\arrayrulewidth}
		\textcolor{model4}{$J_{4AF}$} & \textcolor{model4}{0.0759(4)} &   & \multirow{ 2}{*}{\textcolor{model4}{-0.0085(1)}} \\  
		\textcolor{model4}{$J_{4F}$} & \textcolor{model4}{-0.0504(4)}  &   & \\ \Xhline{3\arrayrulewidth}
		$T\rm_{_{N}}$	& 24.4(3)\footnote{Calculated using the mean field estimate $T_{N}$~$\sim$$\left|\frac{2}{3}S(S+1)z_{2}J_{2} \right|$}   & 25.1(4)~\cite{jauch}  &   \\ 
		\Xhline{3\arrayrulewidth}
		$\theta\rm_{_{CW}}$	& -25.4(5)  & -28.4(4)~\cite{weiss,singer} &  \\ 
		%\Xhline{3\arrayrulewidth}
	\end{tabular}
	\end{ruledtabular}
\label{tab:tab1}
\end{table*}

While the excitation energy provides the magnitude $|J|$, the neutron spectroscopic momentum dependence can be used to extract the corresponding intra-pair distance $\mathbf{R}_{m}$, where $m$ denotes the coordination shell.  By applying the Hohenberg-Brinckman first moment sum rule and the single mode approximation for an isolated pair, excitations from a Co$^{2+}$ pair have the following $|\mathbf{Q|}$ dependence~\cite{dimer,sumrules,stone}
\begin{equation}
S(|\mathbf{Q}|) \propto \frac{|F(|\mathbf{Q}|)|^{2}}{\Delta E\rm_{_{o}}} \left(1-\frac{\sin(|\mathbf{Q}||\mathbf{R}_{m}|)}{|\mathbf{Q}||\mathbf{R}_{m}|} \right),
\label{eq:4}
\end{equation}
\noindent  with $|F(|\mathbf{Q}|)|^{2}$ the magnetic form factor.  Since the modulation is solely dependent on the intra-pair distance $\mathbf{R}_{m}$, the excitation can be assigned to a particular pair and corresponding coordination shell in the $Fm\bar{3}m$ structure as illustrated in Fig.~\ref{fig:fig1}$(a)$. 

\section{Results and Discussion:}

\indent Having discussed the theory for isolated pairs in dilute \mgcoo, we now present the experimental data.  As illustrated by Fig.~\ref{fig:fig2}$(a)$, low temperature/incident energy inelastic neutron spectroscopic measurements on powder \mgcoo~ display a hierarchy of dispersionless excitations up to $\Delta E \sim$ 15~meV.  Based on the energy value of the excitations, we can assign an exchange constant as shown in Fig.~\ref{fig:fig2}$(b)$ using the previously measured value for the spin-orbit coupling constant $\widetilde{\lambda}$~\cite{cowley} for isolated Co$^{2+}$ on a rocksalt lattice. The intensities for each of the seven excitations in Fig.~\ref{fig:fig2}$(a)$ exhibit a modulated \QQ-dependence, characteristic of pairwise interactions and thus distinguishing them from single-ion dispersionless crystal field excitations~\cite{dimer}.  As shown in Figs.~\ref{fig:fig3}$(a,b)$, by fitting the intensity of each mode at different energies to Eq.~\ref{eq:4}, the different pair excitations could be assigned to relative coordination shells ranging from $m$=1 to $m$=4.

We now discuss the temperature dependence with the goal of extracting the sign of $J$.  Antiferromagnetically coupled ($J>$~0) pairs of $j\rm_{_{eff}}=\frac{1}{2}$ spins consist of a singlet ground state and a triplet excited state while ferromagnetic coupling ($J<$~0) gives a triplet ground state and a single excited state.  These two different coupling scenarios give distinct temperature dependences of the integrated intensity that scales as the thermal population difference between the ground and excited states~\cite{zhu,stone}, with antiferromagnetic pairs following 

\begin{equation}
I_{AF}(T)~\propto~(1-e^{-\Delta E/k_{B}T})/(1+3e^{-\Delta E/k_{B}T})
\end{equation}

\noindent and ferromagnetic pairs 

\begin{equation}
I_{F}(T)\propto (1-e^{-\Delta E/k_{B}T})/(3+e^{-\Delta E/k_{B}T}),
\end{equation}

\noindent such that as $T \rightarrow 0$~K, the ratio 

\begin{equation}
\frac{I_{AF}}{I_{F}}~=~\frac{3+e^{-\Delta E/k_{B}T}}{1+3e^{-\Delta E/k_{B}T}}\rightarrow3.
\end{equation}

\noindent As illustrated in Fig.~\ref{fig:fig3}$(d)$, by normalizing the temperature dependence by $I_{F}(T)$, all integrated intensities fall onto either one of two universal curves describing antiferromagnetism or ferromagnetism.

\indent All extracted values of $J$ based on the energy, momentum, and temperature dependence discussed above are summarized in Tab.~\ref{tab:tab1}.   All coordination shells, with the exception of $m$~=~2, display two closely spaced excitations with differing signs for the exchange constant as illustrated in Fig.~\ref{fig:fig3}$(c)$ for the $\sim$ 5 meV excitation.   This presence of dual ferro and antiferromagnetic interactions for $m$~=~1, 3 and 4 is consistent with the GKA rules~\cite{rules1,rules2,rules3,rules4} since each of these exchange pathways, consists of at least one 90$\rm{^{o}}$ Co$^{2+}$-Co$^{2+}$ interaction involving the overlap of half and filled orbitals.  Indeed, the GKA rules predict that the combination of the orbital degree of freedom for \emph{each} Co$^{2+}$ and a lack of orbital ordering (or anisotropy) would manifest itself as either a direct antiferromagnetic $t\rm{_{2g}}^{1}$-$t\rm{_{2g}}^{1}$ or a weaker ferromagnetic $t\rm{_{2g}}^{1}$-$t\rm{_{2g}}^{2}$ exchange interaction.  As summarized in Fig.~\ref{fig:fig3}$(d)$ and Tab.~\ref{tab:tab1}, the experimental results verify the GKA rules~\cite{rules1,rules2,rules3,rules4} as the antiferromagnetic interaction is stronger than the ferromagnetic alternative for all the $m~\neq~2$ excitations while the 180$\rm{^{o}}$ Co$^{2+}$-O$^{2-}$-Co$^{2+}$ $m$~=~2 coupling leads to only a strong antiferromagnetic interaction. 

\indent Having assigned the signs of the 7 exchange constants for dilute Co$_{0.03}$Mg$_{0.97}$O, we now provide a comparison with thermodynamic data and previously measured and calculated exchange constants for bulk CoO.  The additional complication of a dual ferro and antiferromagnetic interactions for most $m$ exchange pathways in combination to the entanglement of individual spin-orbit manifolds in the presence of magnetic order provides a possible explanation for the large range of $J$ values reported for CoO~\cite{2011,tomiyasu,1957,att2,att3,att4,att5,satoh,kant}.  As summarized in Tab.~\ref{tab:tab1}, the values of $J$  show good agreement with three general trends reported by experiment~\cite{2011}: (i) dominant $J_{2}$~$>$~0, (ii) a $J_{1}$~$<$~0 and (iii) a significantly smaller but non-negligible $J_{3}$, all in broad agreement with the trends concluded from a recent $GGA+U$ DFT calculation on CoO (though no such dual exchange was predicted)~\cite{DFT}. In terms of thermodynamic data, the Curie-Weiss constant is related to the exchange interactions $via$ $\Theta_{CW}=-\frac{2}{3} S(S+1) \sum_{i}z_{i}J_{i}$, where the spin value $S=\frac{3}{2}$ and $z_{i}$ is the number of neighbors for each $i^{\rm_{th}}$ exchange interaction~\cite{MFtheory,kittel}.  Following Kanamori~\cite{1957} and applying a correction for spin-orbit coupling, the effective Curie-Weiss temperature $\widetilde{\theta}\rm_{_{CW}}$ is listed in Tab.~\ref{tab:tab1} and compared against a mean field $T_{N}$ calculated based just on $J_{2}$. The estimated $\widetilde{\theta}\rm_{_{CW}}$ of $-295(5)$~K ($-25.4(5)$~meV) and a mean-field estimate of T$\rm_{_{N}}$ of 283(5)~K (24.4(3)~meV), demonstrate close similarities with experimentally determined values of $\theta\rm{_{CW}}$~=~$-330 (4)$~K~\cite{weiss,singer} and T$\rm_{_{N}}$~=~291(4)~K~\cite{jauch}, respectively, for CoO.  The excellent agreement results from the near perfect cancellation of antiferromagnetic and ferromagnetic interactions for all coordinations with the exception of $m$~=~2 (the 180$^{\circ}$ interaction). Although the Co$_{0.03}$Mg$_{0.97}$O lattice ($a$~=~4.21~\AA) is contracted relative to that of pure CoO ($a$~=~4.26~\AA~\cite{lattice_parameter_Co}), the above agreements of energy scale are highly suggestive that the Co$^{2+}$-Co$^{2+}$ exchange interactions are not greatly changed, or at least any changes are smaller than  systematic errors introduced by attempting to simplify the scheme in pure CoO. Hence the present results represent a comprehensive set of interaction energy estimates for CoO. 

\indent In summary, we have disentangled the exchange and spin-orbit interactions for Co$^{2+}$ on a rocksalt lattice.  Through a combined analysis of the energy, momentum, and temperature dependence, we have extracted 7 exchange constants out to four coordination shells. Both ferro and antiferromagnetic interactions are observed with the exception of second neighbor interactions through linear Co$^{2+}$-O$^{2-}$-Co$^{2+}$ bridges, in agreement with both the GKA rules and thermodynamic data.  The results demonstrate that in the case of an orbital degeneracy in the $t_{2g}$ channel dual ferro and antiferromagnetic interactions occur with comparable magnitudes. 

\indent We acknowledge useful conversations with T.~Guidi, J.~R.~Stewart, M.~A.~Green, T.~J.~Williams, K.~H.~Hong, G.~M. McNally and S.~E.~Maytham. We are grateful to the Carnegie Trust for the Universities of Scotland, the Royal Society, the STFC, the ERC and the EPSRC for financial support. P.M.S. acknowledges financial support from the CCSF and the University of Edinburgh through the GRS and PCDS.

\renewcommand{\figurename}{Figure S}
\renewcommand{\topfraction}{0.85}
\renewcommand{\bottomfraction}{0.85}
\renewcommand{\textfraction}{0.1}
\renewcommand{\floatpagefraction}{0.75}

%\bibliography{Supplementary_Material}
%\bibliography{references_Co}

\begin{thebibliography}{59}%
\makeatletter
\providecommand \@ifxundefined [1]{%
 \@ifx{#1\undefined}
}%
\providecommand \@ifnum [1]{%
 \ifnum #1\expandafter \@firstoftwo
 \else \expandafter \@secondoftwo
 \fi
}%
\providecommand \@ifx [1]{%
 \ifx #1\expandafter \@firstoftwo
 \else \expandafter \@secondoftwo
 \fi
}%
\providecommand \natexlab [1]{#1}%
\providecommand \enquote  [1]{``#1''}%
\providecommand \bibnamefont  [1]{#1}%
\providecommand \bibfnamefont [1]{#1}%
\providecommand \citenamefont [1]{#1}%
\providecommand \href@noop [0]{\@secondoftwo}%
\providecommand \href [0]{\begingroup \@sanitize@url \@href}%
\providecommand \@href[1]{\@@startlink{#1}\@@href}%
\providecommand \@@href[1]{\endgroup#1\@@endlink}%
\providecommand \@sanitize@url [0]{\catcode `\\12\catcode `\$12\catcode
  `\&12\catcode `\#12\catcode `\^12\catcode `\_12\catcode `\%12\relax}%
\providecommand \@@startlink[1]{}%
\providecommand \@@endlink[0]{}%
\providecommand \url  [0]{\begingroup\@sanitize@url \@url }%
\providecommand \@url [1]{\endgroup\@href {#1}{\urlprefix }}%
\providecommand \urlprefix  [0]{URL }%
\providecommand \Eprint [0]{\href }%
\providecommand \doibase [0]{http://dx.doi.org/}%
\providecommand \selectlanguage [0]{\@gobble}%
\providecommand \bibinfo  [0]{\@secondoftwo}%
\providecommand \bibfield  [0]{\@secondoftwo}%
\providecommand \translation [1]{[#1]}%
\providecommand \BibitemOpen [0]{}%
\providecommand \bibitemStop [0]{}%
\providecommand \bibitemNoStop [0]{.\EOS\space}%
\providecommand \EOS [0]{\spacefactor3000\relax}%
\providecommand \BibitemShut  [1]{\csname bibitem#1\endcsname}%
\let\auto@bib@innerbib\@empty
%</preamble>
\bibitem [{\citenamefont {Tokura}\ and\ \citenamefont
  {Nagaosa}(2000)}]{Tokura}%
  \BibitemOpen
  \bibfield  {author} {\bibinfo {author} {\bibfnamefont {Y.}~\bibnamefont
  {Tokura}}\ and\ \bibinfo {author} {\bibfnamefont {N.}~\bibnamefont
  {Nagaosa}},\ }\href {\doibase 10.1126/science.288.5465.462} {\bibfield
  {journal} {\bibinfo  {journal} {Science}\ }\textbf {\bibinfo {volume}
  {288}},\ \bibinfo {pages} {462} (\bibinfo {year} {2000})}\BibitemShut
  {NoStop}%
\bibitem [{\citenamefont {Dagotto}(2005)}]{Dagotto}%
  \BibitemOpen
  \bibfield  {author} {\bibinfo {author} {\bibfnamefont {E.}~\bibnamefont
  {Dagotto}},\ }\href {\doibase 10.1126/science.1107559} {\bibfield  {journal}
  {\bibinfo  {journal} {Science}\ }\textbf {\bibinfo {volume} {309}},\ \bibinfo
  {pages} {257} (\bibinfo {year} {2005})}\BibitemShut {NoStop}%
\bibitem [{\citenamefont {Kugel}\ and\ \citenamefont {Khomskii}(1982)}]{Kugel}%
  \BibitemOpen
  \bibfield  {author} {\bibinfo {author} {\bibfnamefont {K.~I.}\ \bibnamefont
  {Kugel}}\ and\ \bibinfo {author} {\bibfnamefont {D.~I.}\ \bibnamefont
  {Khomskii}},\ }\href@noop {} {\bibfield  {journal} {\bibinfo  {journal}
  {Phys. Usp.}\ }\textbf {\bibinfo {volume} {136}},\ \bibinfo {pages} {621}
  (\bibinfo {year} {1982})}\BibitemShut {NoStop}%
\bibitem [{\citenamefont {Okamoto}\ \emph {et~al.}(2007)\citenamefont
  {Okamoto}, \citenamefont {Nohara}, \citenamefont {Katori},\ and\
  \citenamefont {Takagi}}]{Okamoto07}%
  \BibitemOpen
  \bibfield  {author} {\bibinfo {author} {\bibfnamefont {Y.}~\bibnamefont
  {Okamoto}}, \bibinfo {author} {\bibfnamefont {M.}~\bibnamefont {Nohara}},
  \bibinfo {author} {\bibfnamefont {H.~A.}\ \bibnamefont {Katori}}, \ and\
  \bibinfo {author} {\bibfnamefont {H.}~\bibnamefont {Takagi}},\ }\href
  {\doibase 10.1103/PhysRevLett.99.137207} {\bibfield  {journal} {\bibinfo
  {journal} {Phys. Rev. Lett.}\ }\textbf {\bibinfo {volume} {99}},\ \bibinfo
  {pages} {137207} (\bibinfo {year} {2007})}\BibitemShut {NoStop}%
\bibitem [{\citenamefont {Wang}\ \emph {et~al.}(2017)\citenamefont {Wang},
  \citenamefont {Go},\ and\ \citenamefont {Millis}}]{Wang17}%
  \BibitemOpen
  \bibfield  {author} {\bibinfo {author} {\bibfnamefont {R.}~\bibnamefont
  {Wang}}, \bibinfo {author} {\bibfnamefont {A.}~\bibnamefont {Go}}, \ and\
  \bibinfo {author} {\bibfnamefont {A.~J.}\ \bibnamefont {Millis}},\ }\href
  {\doibase 10.1103/PhysRevB.95.045133} {\bibfield  {journal} {\bibinfo
  {journal} {Phys. Rev. B}\ }\textbf {\bibinfo {volume} {95}},\ \bibinfo
  {pages} {045133} (\bibinfo {year} {2017})}\BibitemShut {NoStop}%
\bibitem [{\citenamefont {Jackeli}\ and\ \citenamefont
  {Khaliullin}(2009)}]{Jackeli09}%
  \BibitemOpen
  \bibfield  {author} {\bibinfo {author} {\bibfnamefont {G.}~\bibnamefont
  {Jackeli}}\ and\ \bibinfo {author} {\bibfnamefont {G.}~\bibnamefont
  {Khaliullin}},\ }\href {\doibase 10.1103/PhysRevLett.102.017205} {\bibfield
  {journal} {\bibinfo  {journal} {Phys. Rev. Lett.}\ }\textbf {\bibinfo
  {volume} {102}},\ \bibinfo {pages} {017205} (\bibinfo {year}
  {2009})}\BibitemShut {NoStop}%
\bibitem [{\citenamefont {Shull}\ \emph {et~al.}(1951)\citenamefont {Shull},
  \citenamefont {Strauser},\ and\ \citenamefont {Wollan}}]{Shull1951}%
  \BibitemOpen
  \bibfield  {author} {\bibinfo {author} {\bibfnamefont {C.~G.}\ \bibnamefont
  {Shull}}, \bibinfo {author} {\bibfnamefont {W.~A.}\ \bibnamefont {Strauser}},
  \ and\ \bibinfo {author} {\bibfnamefont {E.~O.}\ \bibnamefont {Wollan}},\
  }\href {\doibase 10.1103/PhysRev.83.333} {\bibfield  {journal} {\bibinfo
  {journal} {Phys. Rev.}\ }\textbf {\bibinfo {volume} {83}},\ \bibinfo {pages}
  {333} (\bibinfo {year} {1951})}\BibitemShut {NoStop}%
\bibitem [{\citenamefont {Li}(1955)}]{Li1955}%
  \BibitemOpen
  \bibfield  {author} {\bibinfo {author} {\bibfnamefont {Y.~Y.}\ \bibnamefont
  {Li}},\ }\href {\doibase 10.1103/PhysRev.100.627} {\bibfield  {journal}
  {\bibinfo  {journal} {Phys. Rev.}\ }\textbf {\bibinfo {volume} {100}},\
  \bibinfo {pages} {627} (\bibinfo {year} {1955})}\BibitemShut {NoStop}%
\bibitem [{\citenamefont {Roth}(1958)}]{structure1}%
  \BibitemOpen
  \bibfield  {author} {\bibinfo {author} {\bibfnamefont {W.~L.}\ \bibnamefont
  {Roth}},\ }\href {\doibase 10.1103/PhysRev.110.1333} {\bibfield  {journal}
  {\bibinfo  {journal} {Phys. Rev.}\ }\textbf {\bibinfo {volume} {110}},\
  \bibinfo {pages} {1333} (\bibinfo {year} {1958})}\BibitemShut {NoStop}%
\bibitem [{\citenamefont {van Laar}(1965)}]{structure2}%
  \BibitemOpen
  \bibfield  {author} {\bibinfo {author} {\bibfnamefont {B.}~\bibnamefont {van
  Laar}},\ }\href {\doibase 10.1103/PhysRev.138.A584} {\bibfield  {journal}
  {\bibinfo  {journal} {Phys. Rev.}\ }\textbf {\bibinfo {volume} {138}},\
  \bibinfo {pages} {A584} (\bibinfo {year} {1965})}\BibitemShut {NoStop}%
\bibitem [{\citenamefont {Satija}\ \emph {et~al.}(1980)\citenamefont {Satija},
  \citenamefont {Axe}, \citenamefont {Shirane}, \citenamefont {Yoshizawa},\
  and\ \citenamefont {Hirakawa}}]{Satija}%
  \BibitemOpen
  \bibfield  {author} {\bibinfo {author} {\bibfnamefont {S.~K.}\ \bibnamefont
  {Satija}}, \bibinfo {author} {\bibfnamefont {J.~D.}\ \bibnamefont {Axe}},
  \bibinfo {author} {\bibfnamefont {G.}~\bibnamefont {Shirane}}, \bibinfo
  {author} {\bibfnamefont {H.}~\bibnamefont {Yoshizawa}}, \ and\ \bibinfo
  {author} {\bibfnamefont {K.}~\bibnamefont {Hirakawa}},\ }\href {\doibase
  10.1103/PhysRevB.21.2001} {\bibfield  {journal} {\bibinfo  {journal} {Phys.
  Rev. B}\ }\textbf {\bibinfo {volume} {21}},\ \bibinfo {pages} {2001}
  (\bibinfo {year} {1980})}\BibitemShut {NoStop}%
\bibitem [{\citenamefont {Oles}\ \emph {et~al.}(2006)\citenamefont {Oles},
  \citenamefont {Horsch}, \citenamefont {Feiner},\ and\ \citenamefont
  {Khaliullin}}]{Oles}%
  \BibitemOpen
  \bibfield  {author} {\bibinfo {author} {\bibfnamefont {A.~M.}\ \bibnamefont
  {Oles}}, \bibinfo {author} {\bibfnamefont {P.}~\bibnamefont {Horsch}},
  \bibinfo {author} {\bibfnamefont {L.~F.}\ \bibnamefont {Feiner}}, \ and\
  \bibinfo {author} {\bibfnamefont {G.}~\bibnamefont {Khaliullin}},\ }\href
  {\doibase 10.1103/PhysRevLett.96.147205} {\bibfield  {journal} {\bibinfo
  {journal} {Phys. Rev. Lett.}\ }\textbf {\bibinfo {volume} {96}},\ \bibinfo
  {pages} {147205} (\bibinfo {year} {2006})}\BibitemShut {NoStop}%
\bibitem [{\citenamefont {Haverkort}\ \emph {et~al.}(2007)\citenamefont
  {Haverkort}, \citenamefont {Tanaka}, \citenamefont {Tjeng},\ and\
  \citenamefont {Sawatzky}}]{haverkort2007}%
  \BibitemOpen
  \bibfield  {author} {\bibinfo {author} {\bibfnamefont {M.~W.}\ \bibnamefont
  {Haverkort}}, \bibinfo {author} {\bibfnamefont {A.}~\bibnamefont {Tanaka}},
  \bibinfo {author} {\bibfnamefont {L.~H.}\ \bibnamefont {Tjeng}}, \ and\
  \bibinfo {author} {\bibfnamefont {G.~A.}\ \bibnamefont {Sawatzky}},\ }\href
  {\doibase 10.1103/PhysRevLett.99.257401} {\bibfield  {journal} {\bibinfo
  {journal} {Physical Rev. Lett.}\ }\textbf {\bibinfo {volume} {99}},\ \bibinfo
  {pages} {257401} (\bibinfo {year} {2007})}\BibitemShut {NoStop}%
\bibitem [{\citenamefont {Larson}\ \emph {et~al.}(2007)\citenamefont {Larson},
  \citenamefont {Ku}, \citenamefont {Tischler}, \citenamefont {Lee},
  \citenamefont {Restrepo}, \citenamefont {Eguiluz}, \citenamefont {Zschack},\
  and\ \citenamefont {Finkelstein}}]{larson2007}%
  \BibitemOpen
  \bibfield  {author} {\bibinfo {author} {\bibfnamefont {B.~C.}\ \bibnamefont
  {Larson}}, \bibinfo {author} {\bibfnamefont {W.}~\bibnamefont {Ku}}, \bibinfo
  {author} {\bibfnamefont {J.~Z.}\ \bibnamefont {Tischler}}, \bibinfo {author}
  {\bibfnamefont {C.-C.}\ \bibnamefont {Lee}}, \bibinfo {author} {\bibfnamefont
  {O.~D.}\ \bibnamefont {Restrepo}}, \bibinfo {author} {\bibfnamefont {A.~G.}\
  \bibnamefont {Eguiluz}}, \bibinfo {author} {\bibfnamefont {P.}~\bibnamefont
  {Zschack}}, \ and\ \bibinfo {author} {\bibfnamefont {K.~D.}\ \bibnamefont
  {Finkelstein}},\ }\href {\doibase 10.1103/PhysRevLett.99.026401} {\bibfield
  {journal} {\bibinfo  {journal} {Phys. Rev. Lett.}\ }\textbf {\bibinfo
  {volume} {99}},\ \bibinfo {pages} {026401} (\bibinfo {year}
  {2007})}\BibitemShut {NoStop}%
\bibitem [{\citenamefont {Sakurai}\ \emph {et~al.}(1968)\citenamefont
  {Sakurai}, \citenamefont {Buyers}, \citenamefont {Cowley},\ and\
  \citenamefont {Dolling}}]{sakurai}%
  \BibitemOpen
  \bibfield  {author} {\bibinfo {author} {\bibfnamefont {J.}~\bibnamefont
  {Sakurai}}, \bibinfo {author} {\bibfnamefont {W.~J.~L.}\ \bibnamefont
  {Buyers}}, \bibinfo {author} {\bibfnamefont {R.~A.}\ \bibnamefont {Cowley}},
  \ and\ \bibinfo {author} {\bibfnamefont {G.}~\bibnamefont {Dolling}},\ }\href
  {\doibase 10.1103/PhysRev.167.510} {\bibfield  {journal} {\bibinfo  {journal}
  {Phys. Rev.}\ }\textbf {\bibinfo {volume} {167}},\ \bibinfo {pages} {510}
  (\bibinfo {year} {1968})}\BibitemShut {NoStop}%
\bibitem [{\citenamefont {Kanamori}(1957{\natexlab{a}})}]{1957}%
  \BibitemOpen
  \bibfield  {author} {\bibinfo {author} {\bibfnamefont {J.}~\bibnamefont
  {Kanamori}},\ }\href {\doibase https://doi.org/10.1143/PTP.17.177} {\bibfield
   {journal} {\bibinfo  {journal} {Progr. Theor. Phys.}\ }\textbf {\bibinfo
  {volume} {17}},\ \bibinfo {pages} {177} (\bibinfo {year}
  {1957}{\natexlab{a}})}\BibitemShut {NoStop}%
\bibitem [{\citenamefont {Cowley}\ \emph {et~al.}(2013)\citenamefont {Cowley},
  \citenamefont {Buyers}, \citenamefont {Stock}, \citenamefont {Yamani},
  \citenamefont {Frost}, \citenamefont {Taylor},\ and\ \citenamefont
  {Prabhakaran}}]{cowley}%
  \BibitemOpen
  \bibfield  {author} {\bibinfo {author} {\bibfnamefont {R.~A.}\ \bibnamefont
  {Cowley}}, \bibinfo {author} {\bibfnamefont {W.~J.~L.}\ \bibnamefont
  {Buyers}}, \bibinfo {author} {\bibfnamefont {C.}~\bibnamefont {Stock}},
  \bibinfo {author} {\bibfnamefont {Z.}~\bibnamefont {Yamani}}, \bibinfo
  {author} {\bibfnamefont {C.}~\bibnamefont {Frost}}, \bibinfo {author}
  {\bibfnamefont {J.~W.}\ \bibnamefont {Taylor}}, \ and\ \bibinfo {author}
  {\bibfnamefont {D.}~\bibnamefont {Prabhakaran}},\ }\href {\doibase
  10.1103/PhysRevB.88.205117} {\bibfield  {journal} {\bibinfo  {journal} {Phys.
  Rev. B}\ }\textbf {\bibinfo {volume} {88}},\ \bibinfo {pages} {205117}
  (\bibinfo {year} {2013})}\BibitemShut {NoStop}%
\bibitem [{\citenamefont {Wallington}\ \emph {et~al.}(2015)\citenamefont
  {Wallington}, \citenamefont {Ar\'{e}valo-Lopez}, \citenamefont {Taylor},
  \citenamefont {Stewart}, \citenamefont {Garc\'{i}a-Sakai}, \citenamefont
  {Attfield},\ and\ \citenamefont {Stock}}]{cov2o6}%
  \BibitemOpen
  \bibfield  {author} {\bibinfo {author} {\bibfnamefont {F.}~\bibnamefont
  {Wallington}}, \bibinfo {author} {\bibfnamefont {A.~M.}\ \bibnamefont
  {Ar\'{e}valo-Lopez}}, \bibinfo {author} {\bibfnamefont {J.~W.}\ \bibnamefont
  {Taylor}}, \bibinfo {author} {\bibfnamefont {J.~R.}\ \bibnamefont {Stewart}},
  \bibinfo {author} {\bibfnamefont {V.}~\bibnamefont {Garc\'{i}a-Sakai}},
  \bibinfo {author} {\bibfnamefont {J.~P.}\ \bibnamefont {Attfield}}, \ and\
  \bibinfo {author} {\bibfnamefont {C.}~\bibnamefont {Stock}},\ }\href
  {\doibase 10.1103/PhysRevB.92.125116} {\bibfield  {journal} {\bibinfo
  {journal} {Phys. Rev. B}\ }\textbf {\bibinfo {volume} {92}},\ \bibinfo
  {pages} {125116} (\bibinfo {year} {2015})}\BibitemShut {NoStop}%
\bibitem [{\citenamefont {Khomskii}(2014)}]{projection}%
  \BibitemOpen
  \bibfield  {author} {\bibinfo {author} {\bibfnamefont {D.~I.}\ \bibnamefont
  {Khomskii}},\ }\href {\doibase 10.1017/CBO9781139096782} {\emph {\bibinfo
  {title} {{Transition Metal Compounds}}}}\ (\bibinfo  {publisher} {Cambridge
  University Press},\ \bibinfo {year} {2014})\BibitemShut {NoStop}%
\bibitem [{\citenamefont {Abragam}\ and\ \citenamefont
  {Bleaney}(2012)}]{projection2}%
  \BibitemOpen
  \bibfield  {author} {\bibinfo {author} {\bibfnamefont {A.}~\bibnamefont
  {Abragam}}\ and\ \bibinfo {author} {\bibfnamefont {B.}~\bibnamefont
  {Bleaney}},\ }\href@noop {} {\emph {\bibinfo {title} {{Electron paramagnetic
  resonance of transition ions}}}}\ (\bibinfo  {publisher} {OUP Oxford},\
  \bibinfo {year} {2012})\BibitemShut {NoStop}%
\bibitem [{\citenamefont {Kanamori}(1957{\natexlab{b}})}]{1957_2}%
  \BibitemOpen
  \bibfield  {author} {\bibinfo {author} {\bibfnamefont {J.}~\bibnamefont
  {Kanamori}},\ }\href {\doibase https://doi.org/10.1143/PTP.17.197} {\bibfield
   {journal} {\bibinfo  {journal} {Progr. Theor. Phys.}\ }\textbf {\bibinfo
  {volume} {17}},\ \bibinfo {pages} {197} (\bibinfo {year}
  {1957}{\natexlab{b}})}\BibitemShut {NoStop}%
\bibitem [{\citenamefont {Buyers}\ \emph {et~al.}(1971)\citenamefont {Buyers},
  \citenamefont {Holden}, \citenamefont {Svensson}, \citenamefont {Cowley},\
  and\ \citenamefont {Hutchings}}]{buyers}%
  \BibitemOpen
  \bibfield  {author} {\bibinfo {author} {\bibfnamefont {W.~J.~L.}\
  \bibnamefont {Buyers}}, \bibinfo {author} {\bibfnamefont {T.~M.}\
  \bibnamefont {Holden}}, \bibinfo {author} {\bibfnamefont {E.~C.}\
  \bibnamefont {Svensson}}, \bibinfo {author} {\bibfnamefont {R.~A.}\
  \bibnamefont {Cowley}}, \ and\ \bibinfo {author} {\bibfnamefont {M.~T.}\
  \bibnamefont {Hutchings}},\ }\href {\doibase
  https://doi.org/10.1088/0022-3719/4/14/028} {\bibfield  {journal} {\bibinfo
  {journal} {J. Phys. C Solid State Phys.}\ }\textbf {\bibinfo {volume} {4}},\
  \bibinfo {pages} {2139} (\bibinfo {year} {1971})}\BibitemShut {NoStop}%
\bibitem [{\citenamefont {Buyers}\ \emph {et~al.}(1984)\citenamefont {Buyers},
  \citenamefont {Holden}, \citenamefont {Svensson},\ and\ \citenamefont
  {Lockwood}}]{buyers1984}%
  \BibitemOpen
  \bibfield  {author} {\bibinfo {author} {\bibfnamefont {W.~J.~L.}\
  \bibnamefont {Buyers}}, \bibinfo {author} {\bibfnamefont {T.~M.}\
  \bibnamefont {Holden}}, \bibinfo {author} {\bibfnamefont {E.~C.}\
  \bibnamefont {Svensson}}, \ and\ \bibinfo {author} {\bibfnamefont {D.~J.}\
  \bibnamefont {Lockwood}},\ }\href {\doibase 10.1103/PhysRevB.30.6521}
  {\bibfield  {journal} {\bibinfo  {journal} {Phys. Rev. B}\ }\textbf {\bibinfo
  {volume} {30}},\ \bibinfo {pages} {6521} (\bibinfo {year}
  {1984})}\BibitemShut {NoStop}%
\bibitem [{\citenamefont {Cowley}\ \emph {et~al.}(1973)\citenamefont {Cowley},
  \citenamefont {Buyers}, \citenamefont {Martel},\ and\ \citenamefont
  {Stevenson}}]{cowley2}%
  \BibitemOpen
  \bibfield  {author} {\bibinfo {author} {\bibfnamefont {R.~A.}\ \bibnamefont
  {Cowley}}, \bibinfo {author} {\bibfnamefont {W.~J.~L.}\ \bibnamefont
  {Buyers}}, \bibinfo {author} {\bibfnamefont {P.}~\bibnamefont {Martel}}, \
  and\ \bibinfo {author} {\bibfnamefont {R.~W.~H.}\ \bibnamefont {Stevenson}},\
  }\href {\doibase https://doi.org/10.1088/0022-3719/6/20/014} {\bibfield
  {journal} {\bibinfo  {journal} {J. Phys. C Solid State Phys.}\ }\textbf
  {\bibinfo {volume} {6}},\ \bibinfo {pages} {2997} (\bibinfo {year}
  {1973})}\BibitemShut {NoStop}%
\bibitem [{\citenamefont {Jauch}\ \emph {et~al.}(2001)\citenamefont {Jauch},
  \citenamefont {Reehuis}, \citenamefont {Bleif}, \citenamefont {Kubanek},\
  and\ \citenamefont {Pattison}}]{jauch}%
  \BibitemOpen
  \bibfield  {author} {\bibinfo {author} {\bibfnamefont {W.}~\bibnamefont
  {Jauch}}, \bibinfo {author} {\bibfnamefont {M.}~\bibnamefont {Reehuis}},
  \bibinfo {author} {\bibfnamefont {H.~J.}\ \bibnamefont {Bleif}}, \bibinfo
  {author} {\bibfnamefont {F.}~\bibnamefont {Kubanek}}, \ and\ \bibinfo
  {author} {\bibfnamefont {P.}~\bibnamefont {Pattison}},\ }\href {\doibase
  10.1103/PhysRevB.64.052102} {\bibfield  {journal} {\bibinfo  {journal} {Phys.
  Rev. B}\ }\textbf {\bibinfo {volume} {64}},\ \bibinfo {pages} {052102}
  (\bibinfo {year} {2001})}\BibitemShut {NoStop}%
\bibitem [{\citenamefont {Coldea}\ \emph {et~al.}(2010)\citenamefont {Coldea},
  \citenamefont {Tennant}, \citenamefont {Wheeler}, \citenamefont {Wawrzynska},
  \citenamefont {Prabhakaran}, \citenamefont {Telling}, \citenamefont
  {Habicht}, \citenamefont {Smeibidl},\ and\ \citenamefont {Kiefer}}]{coldea}%
  \BibitemOpen
  \bibfield  {author} {\bibinfo {author} {\bibfnamefont {R.}~\bibnamefont
  {Coldea}}, \bibinfo {author} {\bibfnamefont {D.~A.}\ \bibnamefont {Tennant}},
  \bibinfo {author} {\bibfnamefont {E.~M.}\ \bibnamefont {Wheeler}}, \bibinfo
  {author} {\bibfnamefont {E.}~\bibnamefont {Wawrzynska}}, \bibinfo {author}
  {\bibfnamefont {D.}~\bibnamefont {Prabhakaran}}, \bibinfo {author}
  {\bibfnamefont {M.}~\bibnamefont {Telling}}, \bibinfo {author} {\bibfnamefont
  {K.}~\bibnamefont {Habicht}}, \bibinfo {author} {\bibfnamefont
  {P.}~\bibnamefont {Smeibidl}}, \ and\ \bibinfo {author} {\bibfnamefont
  {K.}~\bibnamefont {Kiefer}},\ }\href {\doibase 10.1126/science.1180085}
  {\bibfield  {journal} {\bibinfo  {journal} {Science}\ }\textbf {\bibinfo
  {volume} {327}},\ \bibinfo {pages} {177} (\bibinfo {year}
  {2010})}\BibitemShut {NoStop}%
\bibitem [{\citenamefont {Grenier}\ \emph {et~al.}(2015)\citenamefont
  {Grenier}, \citenamefont {Petit}, \citenamefont {Simonet}, \citenamefont
  {Can\'evet}, \citenamefont {Regnault}, \citenamefont {Raymond}, \citenamefont
  {Canals}, \citenamefont {Berthier},\ and\ \citenamefont {Lejay}}]{grenier}%
  \BibitemOpen
  \bibfield  {author} {\bibinfo {author} {\bibfnamefont {B.}~\bibnamefont
  {Grenier}}, \bibinfo {author} {\bibfnamefont {S.}~\bibnamefont {Petit}},
  \bibinfo {author} {\bibfnamefont {V.}~\bibnamefont {Simonet}}, \bibinfo
  {author} {\bibfnamefont {E.}~\bibnamefont {Can\'evet}}, \bibinfo {author}
  {\bibfnamefont {L.-P.}\ \bibnamefont {Regnault}}, \bibinfo {author}
  {\bibfnamefont {S.}~\bibnamefont {Raymond}}, \bibinfo {author} {\bibfnamefont
  {B.}~\bibnamefont {Canals}}, \bibinfo {author} {\bibfnamefont
  {C.}~\bibnamefont {Berthier}}, \ and\ \bibinfo {author} {\bibfnamefont
  {P.}~\bibnamefont {Lejay}},\ }\href {\doibase 10.1103/PhysRevLett.114.017201}
  {\bibfield  {journal} {\bibinfo  {journal} {Phys. Rev. Lett.}\ }\textbf
  {\bibinfo {volume} {114}},\ \bibinfo {pages} {017201} (\bibinfo {year}
  {2015})}\BibitemShut {NoStop}%
\bibitem [{\citenamefont {Zhou}\ \emph {et~al.}(2012)\citenamefont {Zhou},
  \citenamefont {Xu}, \citenamefont {Hallas}, \citenamefont {Silverstein},
  \citenamefont {Wiebe}, \citenamefont {Umegaki}, \citenamefont {Yan},
  \citenamefont {Murphy}, \citenamefont {Park}, \citenamefont {Qiu},
  \citenamefont {Copley}, \citenamefont {Gardner},\ and\ \citenamefont
  {Takano}}]{zhou2012}%
  \BibitemOpen
  \bibfield  {author} {\bibinfo {author} {\bibfnamefont {H.~D.}\ \bibnamefont
  {Zhou}}, \bibinfo {author} {\bibfnamefont {C.}~\bibnamefont {Xu}}, \bibinfo
  {author} {\bibfnamefont {A.~M.}\ \bibnamefont {Hallas}}, \bibinfo {author}
  {\bibfnamefont {H.~J.}\ \bibnamefont {Silverstein}}, \bibinfo {author}
  {\bibfnamefont {C.~R.}\ \bibnamefont {Wiebe}}, \bibinfo {author}
  {\bibfnamefont {I.}~\bibnamefont {Umegaki}}, \bibinfo {author} {\bibfnamefont
  {J.~Q.}\ \bibnamefont {Yan}}, \bibinfo {author} {\bibfnamefont {T.~P.}\
  \bibnamefont {Murphy}}, \bibinfo {author} {\bibfnamefont {J.-H.}\
  \bibnamefont {Park}}, \bibinfo {author} {\bibfnamefont {Y.}~\bibnamefont
  {Qiu}}, \bibinfo {author} {\bibfnamefont {J.~R.~D.}\ \bibnamefont {Copley}},
  \bibinfo {author} {\bibfnamefont {J.~S.}\ \bibnamefont {Gardner}}, \ and\
  \bibinfo {author} {\bibfnamefont {Y.}~\bibnamefont {Takano}},\ }\href
  {\doibase 10.1103/PhysRevLett.109.267206} {\bibfield  {journal} {\bibinfo
  {journal} {Phys. Rev. Lett.}\ }\textbf {\bibinfo {volume} {109}},\ \bibinfo
  {pages} {267206} (\bibinfo {year} {2012})}\BibitemShut {NoStop}%
\bibitem [{\citenamefont {Dalverny}\ \emph {et~al.}(2010)\citenamefont
  {Dalverny}, \citenamefont {Filhol}, \citenamefont {Lemoigno},\ and\
  \citenamefont {Doublet}}]{Dalverny}%
  \BibitemOpen
  \bibfield  {author} {\bibinfo {author} {\bibfnamefont {A.~L.}\ \bibnamefont
  {Dalverny}}, \bibinfo {author} {\bibfnamefont {J.~S.}\ \bibnamefont
  {Filhol}}, \bibinfo {author} {\bibfnamefont {F.}~\bibnamefont {Lemoigno}}, \
  and\ \bibinfo {author} {\bibfnamefont {M.~L.}\ \bibnamefont {Doublet}},\
  }\href {\doibase 10.1021/jp108599m} {\bibfield  {journal} {\bibinfo
  {journal} {J. Phys. Chem. C}\ }\textbf {\bibinfo {volume} {114}},\ \bibinfo
  {pages} {21750} (\bibinfo {year} {2010})}\BibitemShut {NoStop}%
\bibitem [{\citenamefont {Feygenson}\ \emph {et~al.}(2011)\citenamefont
  {Feygenson}, \citenamefont {Teng}, \citenamefont {Inderhees}, \citenamefont
  {Yiu}, \citenamefont {Du}, \citenamefont {Han}, \citenamefont {Wen},
  \citenamefont {Xu}, \citenamefont {Podlesnyak}, \citenamefont {Niedziela},
  \citenamefont {Hagen}, \citenamefont {Qiu}, \citenamefont {Brown},
  \citenamefont {Zhang},\ and\ \citenamefont {Aronson}}]{2011}%
  \BibitemOpen
  \bibfield  {author} {\bibinfo {author} {\bibfnamefont {M.}~\bibnamefont
  {Feygenson}}, \bibinfo {author} {\bibfnamefont {X.}~\bibnamefont {Teng}},
  \bibinfo {author} {\bibfnamefont {S.~E.}\ \bibnamefont {Inderhees}}, \bibinfo
  {author} {\bibfnamefont {Y.}~\bibnamefont {Yiu}}, \bibinfo {author}
  {\bibfnamefont {W.}~\bibnamefont {Du}}, \bibinfo {author} {\bibfnamefont
  {W.}~\bibnamefont {Han}}, \bibinfo {author} {\bibfnamefont {J.}~\bibnamefont
  {Wen}}, \bibinfo {author} {\bibfnamefont {Z.}~\bibnamefont {Xu}}, \bibinfo
  {author} {\bibfnamefont {A.~A.}\ \bibnamefont {Podlesnyak}}, \bibinfo
  {author} {\bibfnamefont {J.~L.}\ \bibnamefont {Niedziela}}, \bibinfo {author}
  {\bibfnamefont {M.}~\bibnamefont {Hagen}}, \bibinfo {author} {\bibfnamefont
  {Y.}~\bibnamefont {Qiu}}, \bibinfo {author} {\bibfnamefont {C.~M.}\
  \bibnamefont {Brown}}, \bibinfo {author} {\bibfnamefont {L.}~\bibnamefont
  {Zhang}}, \ and\ \bibinfo {author} {\bibfnamefont {M.~C.}\ \bibnamefont
  {Aronson}},\ }\href {\doibase 10.1103/PhysRevB.83.174414} {\bibfield
  {journal} {\bibinfo  {journal} {Phys. Rev. B}\ }\textbf {\bibinfo {volume}
  {83}},\ \bibinfo {pages} {174414} (\bibinfo {year} {2011})}\BibitemShut
  {NoStop}%
\bibitem [{\citenamefont {Tomiyasu}\ and\ \citenamefont
  {Itoh}(2006)}]{tomiyasu}%
  \BibitemOpen
  \bibfield  {author} {\bibinfo {author} {\bibfnamefont {K.}~\bibnamefont
  {Tomiyasu}}\ and\ \bibinfo {author} {\bibfnamefont {S.}~\bibnamefont
  {Itoh}},\ }\href {\doibase https://doi.org/10.1143/JPSJ.75.084708} {\bibfield
   {journal} {\bibinfo  {journal} {J. Phys. Soc. Jpn.}\ }\textbf {\bibinfo
  {volume} {75}},\ \bibinfo {pages} {084708} (\bibinfo {year}
  {2006})}\BibitemShut {NoStop}%
\bibitem [{\citenamefont {Deng}\ \emph {et~al.}(2010)\citenamefont {Deng},
  \citenamefont {Li}, \citenamefont {Li}, \citenamefont {Xia}, \citenamefont
  {Walsh},\ and\ \citenamefont {Wei}}]{DFT}%
  \BibitemOpen
  \bibfield  {author} {\bibinfo {author} {\bibfnamefont {H.-X.}\ \bibnamefont
  {Deng}}, \bibinfo {author} {\bibfnamefont {J.}~\bibnamefont {Li}}, \bibinfo
  {author} {\bibfnamefont {S.-S.}\ \bibnamefont {Li}}, \bibinfo {author}
  {\bibfnamefont {J.-B.}\ \bibnamefont {Xia}}, \bibinfo {author} {\bibfnamefont
  {A.}~\bibnamefont {Walsh}}, \ and\ \bibinfo {author} {\bibfnamefont {S.-H.}\
  \bibnamefont {Wei}},\ }\href {\doibase http://dx.doi.org/10.1063/1.3402772}
  {\bibfield  {journal} {\bibinfo  {journal} {Appl. Phys. Lett.}\ }\textbf
  {\bibinfo {volume} {96}},\ \bibinfo {pages} {162508} (\bibinfo {year}
  {2010})}\BibitemShut {NoStop}%
\bibitem [{\citenamefont {Yamani}\ \emph {et~al.}(2010)\citenamefont {Yamani},
  \citenamefont {Buyers}, \citenamefont {Cowley},\ and\ \citenamefont
  {Prabhakaran}}]{yamani2}%
  \BibitemOpen
  \bibfield  {author} {\bibinfo {author} {\bibfnamefont {Z.}~\bibnamefont
  {Yamani}}, \bibinfo {author} {\bibfnamefont {W.~J.~L.}\ \bibnamefont
  {Buyers}}, \bibinfo {author} {\bibfnamefont {R.~A.}\ \bibnamefont {Cowley}},
  \ and\ \bibinfo {author} {\bibfnamefont {D.}~\bibnamefont {Prabhakaran}},\
  }\href {\doibase https://doi.org/10.1139/P10-021} {\bibfield  {journal}
  {\bibinfo  {journal} {Can. J. Phys.}\ }\textbf {\bibinfo {volume} {88}},\
  \bibinfo {pages} {729} (\bibinfo {year} {2010})}\BibitemShut {NoStop}%
\bibitem [{\citenamefont {Fischer}\ \emph {et~al.}(2009)\citenamefont
  {Fischer}, \citenamefont {D{\"a}ne}, \citenamefont {Ernst}, \citenamefont
  {Bruno}, \citenamefont {L{\"u}eders}, \citenamefont {Szotek}, \citenamefont
  {Temmerman},\ and\ \citenamefont {Hergert}}]{fischer2}%
  \BibitemOpen
  \bibfield  {author} {\bibinfo {author} {\bibfnamefont {G.}~\bibnamefont
  {Fischer}}, \bibinfo {author} {\bibfnamefont {M.}~\bibnamefont {D{\"a}ne}},
  \bibinfo {author} {\bibfnamefont {A.}~\bibnamefont {Ernst}}, \bibinfo
  {author} {\bibfnamefont {P.}~\bibnamefont {Bruno}}, \bibinfo {author}
  {\bibfnamefont {M.}~\bibnamefont {L{\"u}eders}}, \bibinfo {author}
  {\bibfnamefont {Z.}~\bibnamefont {Szotek}}, \bibinfo {author} {\bibfnamefont
  {W.}~\bibnamefont {Temmerman}}, \ and\ \bibinfo {author} {\bibfnamefont
  {W.}~\bibnamefont {Hergert}},\ }\href {\doibase 10.1103/PhysRevB.80.014408}
  {\bibfield  {journal} {\bibinfo  {journal} {Phys. Rev. B}\ }\textbf {\bibinfo
  {volume} {80}},\ \bibinfo {pages} {014408} (\bibinfo {year}
  {2009})}\BibitemShut {NoStop}%
\bibitem [{\citenamefont {Kant}\ \emph {et~al.}(2008)\citenamefont {Kant},
  \citenamefont {Rudolf}, \citenamefont {Schrettle}, \citenamefont {Mayr},
  \citenamefont {Deisenhofer}, \citenamefont {Lunkenheimer}, \citenamefont
  {Eremin},\ and\ \citenamefont {Loidl}}]{kant}%
  \BibitemOpen
  \bibfield  {author} {\bibinfo {author} {\bibfnamefont {C.}~\bibnamefont
  {Kant}}, \bibinfo {author} {\bibfnamefont {T.}~\bibnamefont {Rudolf}},
  \bibinfo {author} {\bibfnamefont {F.}~\bibnamefont {Schrettle}}, \bibinfo
  {author} {\bibfnamefont {F.}~\bibnamefont {Mayr}}, \bibinfo {author}
  {\bibfnamefont {J.}~\bibnamefont {Deisenhofer}}, \bibinfo {author}
  {\bibfnamefont {P.}~\bibnamefont {Lunkenheimer}}, \bibinfo {author}
  {\bibfnamefont {M.~V.}\ \bibnamefont {Eremin}}, \ and\ \bibinfo {author}
  {\bibfnamefont {A.}~\bibnamefont {Loidl}},\ }\href {\doibase
  10.1103/PhysRevB.78.245103} {\bibfield  {journal} {\bibinfo  {journal} {Phys.
  Rev. B}\ }\textbf {\bibinfo {volume} {78}},\ \bibinfo {pages} {245103}
  (\bibinfo {year} {2008})}\BibitemShut {NoStop}%
\bibitem [{sup()}]{suppl}%
  \BibitemOpen
  \href@noop {} {}\bibinfo {note} {See Supplemental Material at [URL to be
  inserted by publisher] for a description of sample
  characterization.}\BibitemShut {Stop}%
\bibitem [{\citenamefont {Furrer}\ \emph {et~al.}(2015)\citenamefont {Furrer},
  \citenamefont {Podlesnyak},\ and\ \citenamefont {Kr{\"a}mer}}]{furrer2}%
  \BibitemOpen
  \bibfield  {author} {\bibinfo {author} {\bibfnamefont {A.}~\bibnamefont
  {Furrer}}, \bibinfo {author} {\bibfnamefont {A.}~\bibnamefont {Podlesnyak}},
  \ and\ \bibinfo {author} {\bibfnamefont {K.~W.}\ \bibnamefont {Kr{\"a}mer}},\
  }\href {\doibase 10.1103/PhysRevB.92.104415} {\bibfield  {journal} {\bibinfo
  {journal} {Phys. Rev. B}\ }\textbf {\bibinfo {volume} {92}},\ \bibinfo
  {pages} {104415} (\bibinfo {year} {2015})}\BibitemShut {NoStop}%
\bibitem [{\citenamefont {Svensson}\ \emph {et~al.}(1978)\citenamefont
  {Svensson}, \citenamefont {Harvey}, \citenamefont {Buyers},\ and\
  \citenamefont {Holden}}]{Svensson}%
  \BibitemOpen
  \bibfield  {author} {\bibinfo {author} {\bibfnamefont {E.~C.}\ \bibnamefont
  {Svensson}}, \bibinfo {author} {\bibfnamefont {M.}~\bibnamefont {Harvey}},
  \bibinfo {author} {\bibfnamefont {W.~J.~L.}\ \bibnamefont {Buyers}}, \ and\
  \bibinfo {author} {\bibfnamefont {T.~M.}\ \bibnamefont {Holden}},\ }\href
  {\doibase 10.1063/1.324714} {\bibfield  {journal} {\bibinfo  {journal} {J.
  Appl. Phys.}\ }\textbf {\bibinfo {volume} {49}},\ \bibinfo {pages} {2150}
  (\bibinfo {year} {1978})}\BibitemShut {NoStop}%
\bibitem [{\citenamefont {Vegard}(1921)}]{vegard}%
  \BibitemOpen
  \bibfield  {author} {\bibinfo {author} {\bibfnamefont {L.}~\bibnamefont
  {Vegard}},\ }\href@noop {} {\bibfield  {journal} {\bibinfo  {journal} {Z.
  Phys.}\ }\textbf {\bibinfo {volume} {5}},\ \bibinfo {pages} {17} (\bibinfo
  {year} {1921})}\BibitemShut {NoStop}%
\bibitem [{\citenamefont {Goodenough}(1958)}]{rules2}%
  \BibitemOpen
  \bibfield  {author} {\bibinfo {author} {\bibfnamefont {J.~B.}\ \bibnamefont
  {Goodenough}},\ }\href {\doibase
  https://doi.org/10.1016/0022-3697(58)90107-0} {\bibfield  {journal} {\bibinfo
   {journal} {J. Phys. Chem. Solids}\ }\textbf {\bibinfo {volume} {6}},\
  \bibinfo {pages} {287} (\bibinfo {year} {1958})}\BibitemShut {NoStop}%
\bibitem [{\citenamefont {Kanamori}(1959)}]{rules3}%
  \BibitemOpen
  \bibfield  {author} {\bibinfo {author} {\bibfnamefont {J.}~\bibnamefont
  {Kanamori}},\ }\href {\doibase https://doi.org/10.1016/0022-3697(59)90061-7}
  {\bibfield  {journal} {\bibinfo  {journal} {J. Phys. Chem. Solids}\ }\textbf
  {\bibinfo {volume} {10}},\ \bibinfo {pages} {87} (\bibinfo {year}
  {1959})}\BibitemShut {NoStop}%
\bibitem [{\citenamefont {Anderson}(1950)}]{rules4}%
  \BibitemOpen
  \bibfield  {author} {\bibinfo {author} {\bibfnamefont {P.~W.}\ \bibnamefont
  {Anderson}},\ }\href {\doibase 10.1103/PhysRev.79.350} {\bibfield  {journal}
  {\bibinfo  {journal} {Phys. Rev.}\ }\textbf {\bibinfo {volume} {79}},\
  \bibinfo {pages} {350} (\bibinfo {year} {1950})}\BibitemShut {NoStop}%
\bibitem [{\citenamefont {Haraldsen}\ \emph {et~al.}(2005)\citenamefont
  {Haraldsen}, \citenamefont {Barnes},\ and\ \citenamefont {Musfeldt}}]{dimer}%
  \BibitemOpen
  \bibfield  {author} {\bibinfo {author} {\bibfnamefont {J.~T.}\ \bibnamefont
  {Haraldsen}}, \bibinfo {author} {\bibfnamefont {T.}~\bibnamefont {Barnes}}, \
  and\ \bibinfo {author} {\bibfnamefont {J.~L.}\ \bibnamefont {Musfeldt}},\
  }\href {\doibase 10.1103/PhysRevB.71.064403} {\bibfield  {journal} {\bibinfo
  {journal} {Phys. Rev. B}\ }\textbf {\bibinfo {volume} {71}},\ \bibinfo
  {pages} {064403} (\bibinfo {year} {2005})}\BibitemShut {NoStop}%
\bibitem [{\citenamefont {Rose}(1995)}]{projection3}%
  \BibitemOpen
  \bibfield  {author} {\bibinfo {author} {\bibfnamefont {M.}~\bibnamefont
  {Rose}},\ }\href@noop {} {\emph {\bibinfo {title} {{Elementary Theory of
  Angular Momentum}}}},\ Dover {B}ooks on {P}hysics and {C}hemistry\ (\bibinfo
  {publisher} {Dover},\ \bibinfo {year} {1995})\BibitemShut {NoStop}%
\bibitem [{\citenamefont {Furrer}\ and\ \citenamefont
  {Waldmann}(2013)}]{furrer}%
  \BibitemOpen
  \bibfield  {author} {\bibinfo {author} {\bibfnamefont {A.}~\bibnamefont
  {Furrer}}\ and\ \bibinfo {author} {\bibfnamefont {O.}~\bibnamefont
  {Waldmann}},\ }\href {\doibase 10.1103/RevModPhys.85.367} {\bibfield
  {journal} {\bibinfo  {journal} {Rev. Mod. Phys.}\ }\textbf {\bibinfo {volume}
  {85}},\ \bibinfo {pages} {367} (\bibinfo {year} {2013})}\BibitemShut
  {NoStop}%
\bibitem [{\citenamefont {Nagamiya}\ \emph {et~al.}(1955)\citenamefont
  {Nagamiya}, \citenamefont {Yosida},\ and\ \citenamefont {Kubo}}]{weiss}%
  \BibitemOpen
  \bibfield  {author} {\bibinfo {author} {\bibfnamefont {T.}~\bibnamefont
  {Nagamiya}}, \bibinfo {author} {\bibfnamefont {K.}~\bibnamefont {Yosida}}, \
  and\ \bibinfo {author} {\bibfnamefont {R.~F.}\ \bibnamefont {Kubo}},\ }\href
  {\doibase http://dx.doi.org/10.1080/00018735500101154} {\bibfield  {journal}
  {\bibinfo  {journal} {Adv. Phys.}\ }\textbf {\bibinfo {volume} {4}},\
  \bibinfo {pages} {1} (\bibinfo {year} {1955})}\BibitemShut {NoStop}%
\bibitem [{\citenamefont {Singer}(1956)}]{singer}%
  \BibitemOpen
  \bibfield  {author} {\bibinfo {author} {\bibfnamefont {J.~R.}\ \bibnamefont
  {Singer}},\ }\href {\doibase 10.1103/PhysRev.104.929} {\bibfield  {journal}
  {\bibinfo  {journal} {Phys. Rev.}\ }\textbf {\bibinfo {volume} {104}},\
  \bibinfo {pages} {929} (\bibinfo {year} {1956})}\BibitemShut {NoStop}%
\bibitem [{\citenamefont {Lee}\ \emph {et~al.}(2014)\citenamefont {Lee},
  \citenamefont {Lee}, \citenamefont {Lee},\ and\ \citenamefont
  {Whangbo}}]{MFtheory}%
  \BibitemOpen
  \bibfield  {author} {\bibinfo {author} {\bibfnamefont {K.}~\bibnamefont
  {Lee}}, \bibinfo {author} {\bibfnamefont {J.}~\bibnamefont {Lee}}, \bibinfo
  {author} {\bibfnamefont {C.}~\bibnamefont {Lee}}, \ and\ \bibinfo {author}
  {\bibfnamefont {M.}~\bibnamefont {Whangbo}},\ }\href {\doibase
  https://doi.org/10.5012/bkcs.2014.35.5.1277} {\bibfield  {journal} {\bibinfo
  {journal} {Bull. Korean Chem. Soc.}\ }\textbf {\bibinfo {volume} {35}},\
  \bibinfo {pages} {1277} (\bibinfo {year} {2014})}\BibitemShut {NoStop}%
\bibitem [{\citenamefont {Kittel}(2005)}]{kittel}%
  \BibitemOpen
  \bibfield  {author} {\bibinfo {author} {\bibfnamefont {C.}~\bibnamefont
  {Kittel}},\ }\href@noop {} {\emph {\bibinfo {title} {Introduction to solid
  state physics}}}\ (\bibinfo  {publisher} {Wiley, New York},\ \bibinfo {year}
  {2005})\BibitemShut {NoStop}%
\bibitem [{\citenamefont {Hohenberg}\ and\ \citenamefont
  {Brinkman}(1974)}]{sumrules}%
  \BibitemOpen
  \bibfield  {author} {\bibinfo {author} {\bibfnamefont {P.~C.}\ \bibnamefont
  {Hohenberg}}\ and\ \bibinfo {author} {\bibfnamefont {W.~F.}\ \bibnamefont
  {Brinkman}},\ }\href {\doibase 10.1103/PhysRevB.10.128} {\bibfield  {journal}
  {\bibinfo  {journal} {Phys. Rev. B}\ }\textbf {\bibinfo {volume} {10}},\
  \bibinfo {pages} {128} (\bibinfo {year} {1974})}\BibitemShut {NoStop}%
\bibitem [{\citenamefont {Stone}\ \emph {et~al.}(2008)\citenamefont {Stone},
  \citenamefont {Lumsden}, \citenamefont {Chang}, \citenamefont {Samulon},
  \citenamefont {Batista},\ and\ \citenamefont {Fisher}}]{stone}%
  \BibitemOpen
  \bibfield  {author} {\bibinfo {author} {\bibfnamefont {M.~B.}\ \bibnamefont
  {Stone}}, \bibinfo {author} {\bibfnamefont {M.~D.}\ \bibnamefont {Lumsden}},
  \bibinfo {author} {\bibfnamefont {S.}~\bibnamefont {Chang}}, \bibinfo
  {author} {\bibfnamefont {E.~C.}\ \bibnamefont {Samulon}}, \bibinfo {author}
  {\bibfnamefont {C.~D.}\ \bibnamefont {Batista}}, \ and\ \bibinfo {author}
  {\bibfnamefont {I.~R.}\ \bibnamefont {Fisher}},\ }\href {\doibase
  10.1103/PhysRevLett.100.237201} {\bibfield  {journal} {\bibinfo  {journal}
  {Phys. Rev. Lett.}\ }\textbf {\bibinfo {volume} {100}},\ \bibinfo {pages}
  {237201} (\bibinfo {year} {2008})}\BibitemShut {NoStop}%
\bibitem [{\citenamefont {Zhu}(2005)}]{zhu}%
  \BibitemOpen
  \bibfield  {author} {\bibinfo {author} {\bibfnamefont {Y.}~\bibnamefont
  {Zhu}},\ }\href@noop {} {\emph {\bibinfo {title} {Modern techniques for
  characterizing magnetic materials}}}\ (\bibinfo  {publisher} {Springer
  Science \& Business Media},\ \bibinfo {year} {2005})\BibitemShut {NoStop}%
\bibitem [{\citenamefont {Goodenough}(1955)}]{rules1}%
  \BibitemOpen
  \bibfield  {author} {\bibinfo {author} {\bibfnamefont {J.~B.}\ \bibnamefont
  {Goodenough}},\ }\href {\doibase 10.1103/PhysRev.100.564} {\bibfield
  {journal} {\bibinfo  {journal} {Phys. Rev.}\ }\textbf {\bibinfo {volume}
  {100}},\ \bibinfo {pages} {564} (\bibinfo {year} {1955})}\BibitemShut
  {NoStop}%
\bibitem [{\citenamefont {Tachiki}(1964)}]{att2}%
  \BibitemOpen
  \bibfield  {author} {\bibinfo {author} {\bibfnamefont {M.}~\bibnamefont
  {Tachiki}},\ }\href {\doibase https://doi.org/10.1143/JPSJ.19.454} {\bibfield
   {journal} {\bibinfo  {journal} {J. Phys. Soc. Jpn.}\ }\textbf {\bibinfo
  {volume} {19}},\ \bibinfo {pages} {454} (\bibinfo {year} {1964})}\BibitemShut
  {NoStop}%
\bibitem [{\citenamefont {El-Batanouny}(2002)}]{att3}%
  \BibitemOpen
  \bibfield  {author} {\bibinfo {author} {\bibfnamefont {M.}~\bibnamefont
  {El-Batanouny}},\ }\href {\doibase
  https://doi.org/10.1088/0953-8984/14/24/319} {\bibfield  {journal} {\bibinfo
  {journal} {J. Phys.: Condens. Matter}\ }\textbf {\bibinfo {volume} {14}},\
  \bibinfo {pages} {6281} (\bibinfo {year} {2002})}\BibitemShut {NoStop}%
\bibitem [{\citenamefont {Chou}\ and\ \citenamefont {Fan}(1976)}]{att4}%
  \BibitemOpen
  \bibfield  {author} {\bibinfo {author} {\bibfnamefont {H.-h.}\ \bibnamefont
  {Chou}}\ and\ \bibinfo {author} {\bibfnamefont {H.~Y.}\ \bibnamefont {Fan}},\
  }\href {\doibase 10.1103/PhysRevB.13.3924} {\bibfield  {journal} {\bibinfo
  {journal} {Phys. Rev. B}\ }\textbf {\bibinfo {volume} {13}},\ \bibinfo
  {pages} {3924} (\bibinfo {year} {1976})}\BibitemShut {NoStop}%
\bibitem [{\citenamefont {Hayes}\ and\ \citenamefont {Perry}(1973)}]{att5}%
  \BibitemOpen
  \bibfield  {author} {\bibinfo {author} {\bibfnamefont {R.~R.}\ \bibnamefont
  {Hayes}}\ and\ \bibinfo {author} {\bibfnamefont {C.~H.}\ \bibnamefont
  {Perry}},\ }\href {\doibase https://doi.org/10.1016/0038-1098(73)90757-6}
  {\bibfield  {journal} {\bibinfo  {journal} {Solid State Commun.}\ }\textbf
  {\bibinfo {volume} {13}},\ \bibinfo {pages} {1915} (\bibinfo {year}
  {1973})}\BibitemShut {NoStop}%
\bibitem [{\citenamefont {Satoh}\ \emph {et~al.}(2017)\citenamefont {Satoh},
  \citenamefont {Iida}, \citenamefont {Higuchi}, \citenamefont {Fujii},
  \citenamefont {Koreeda}, \citenamefont {Ueda}, \citenamefont {Shimura},
  \citenamefont {Kuroda}, \citenamefont {Butrim},\ and\ \citenamefont
  {Ivanov}}]{satoh}%
  \BibitemOpen
  \bibfield  {author} {\bibinfo {author} {\bibfnamefont {T.}~\bibnamefont
  {Satoh}}, \bibinfo {author} {\bibfnamefont {R.}~\bibnamefont {Iida}},
  \bibinfo {author} {\bibfnamefont {T.}~\bibnamefont {Higuchi}}, \bibinfo
  {author} {\bibfnamefont {Y.}~\bibnamefont {Fujii}}, \bibinfo {author}
  {\bibfnamefont {A.}~\bibnamefont {Koreeda}}, \bibinfo {author} {\bibfnamefont
  {H.}~\bibnamefont {Ueda}}, \bibinfo {author} {\bibfnamefont {T.}~\bibnamefont
  {Shimura}}, \bibinfo {author} {\bibfnamefont {K.}~\bibnamefont {Kuroda}},
  \bibinfo {author} {\bibfnamefont {V.~I.}\ \bibnamefont {Butrim}}, \ and\
  \bibinfo {author} {\bibfnamefont {B.~A.}\ \bibnamefont {Ivanov}},\ }\href
  {\doibase 10.1038/s41467-017-00616-2} {\bibfield  {journal} {\bibinfo
  {journal} {Nat. Commun.}\ }\textbf {\bibinfo {volume} {8}},\ \bibinfo {pages}
  {638} (\bibinfo {year} {2017})}\BibitemShut {NoStop}%
\bibitem [{\citenamefont {Sasaki}\ \emph {et~al.}(1979)\citenamefont {Sasaki},
  \citenamefont {Fujino},\ and\ \citenamefont
  {Tak{\'E}chi}}]{lattice_parameter_Co}%
  \BibitemOpen
  \bibfield  {author} {\bibinfo {author} {\bibfnamefont {S.}~\bibnamefont
  {Sasaki}}, \bibinfo {author} {\bibfnamefont {K.}~\bibnamefont {Fujino}}, \
  and\ \bibinfo {author} {\bibfnamefont {Y.}~\bibnamefont {Tak{\'E}chi}},\
  }\href {\doibase https://doi.org/10.2183/pjab.55.43} {\bibfield  {journal}
  {\bibinfo  {journal} {Proc. Jpn. Acad. Ser. B Phys. Biol. Sci.}\ }\textbf
  {\bibinfo {volume} {55}},\ \bibinfo {pages} {43} (\bibinfo {year}
  {1979})}\BibitemShut {NoStop}%
\end{thebibliography}


\end{document}

\newcommand\bbone{\ensuremath{\mathbbm{1}}}
\newcommand{\ul}{\underline}
\newcommand{\bp}{{\bf p}}
\newcommand{\vl}{v_{_L}}
\newcommand{\vc}{\mathbf}
\newcommand{\be}{\begin{equation}}
\newcommand{\ee}{\end{equation}}
\newcommand{\bk}{{{\bf{k}}}}
\newcommand{\bK}{{{\bf{K}}}}
\newcommand{\cE}{{{\cal E}}}
\newcommand{\bQ}{{{\bf{Q}}}}
\newcommand{\br}{{{\bf{r}}}}
\newcommand{\bg}{{{\bf{g}}}}
\newcommand{\bG}{{{\bf{G}}}}
\newcommand{\hbr}{{\hat{\bf{r}}}}
\newcommand{\bR}{{{\bf{R}}}}
\newcommand{\bq}{{\bf{q}}}
\newcommand{\hx}{{\hat{x}}}
\newcommand{\hy}{{\hat{y}}}
\newcommand{\hd}{{\hat{\delta}}}
\newcommand{\bea}{\begin{eqnarray}}
\newcommand{\eea}{\end{eqnarray}}
\newcommand{\ra}{\rangle}
\newcommand{\la}{\langle}
\renewcommand{\tt}{{\tilde{t}}}
\newcommand{\upa}{\uparrow}
\newcommand{\dna}{\downarrow}
\newcommand{\bS}{{\bf S}}
\newcommand{\vS}{\vec{S}}
\newcommand{\dg}{{\dagger}}
\newcommand{\pdg}{{\phantom\dagger}}
\newcommand{\tphi}{{\tilde\phi}}
\newcommand{\cf}{{\cal F}}
\newcommand{\ca}{{\cal A}}
\renewcommand{\ni}{\noindent}
\newcommand{\ct}{{\cal T}}
\newcommand{\brf}{\bar{F}}
\newcommand{\brg}{\bar{G}}
\newcommand{\jeff}{j_{\rm eff}}

\title{Supplementary information for ``Disentangling orbital and spin exchange interactions for Co$^{2+}$ on a rocksalt lattice"}

\author{P.~M.~Sarte}
\affiliation{School of Chemistry, University of Edinburgh, Edinburgh EH9 3FJ, United Kingdom}
\affiliation{Centre for Science at Extreme Conditions, University of Edinburgh, Edinburgh EH9 3FD, United Kingdom}
\author{R.~A.~Cowley} \altaffiliation{Deceased 27 January 2015}
\affiliation{Department of Physics, Clarendon Laboratory, University of Oxford, Park Road, Oxford, OX1 3PU, United Kingdom}
\author{E.~E.~Rodriguez} 
\affiliation{Department of Chemistry and Biochemistry, University of Maryland, College Park, Maryland 20742, USA} 
\author{E.~Pachoud} 
\affiliation{School of Chemistry, University of Edinburgh, Edinburgh EH9 3FJ, United Kingdom}
\affiliation{Centre for Science at Extreme Conditions, University of Edinburgh, Edinburgh EH9 3FD, United Kingdom}
\author{D.~Le}
\affiliation{ISIS Facility, Rutherford Appleton Laboratory, Chilton, Didcot OX11 0QX, United Kingdom}
\author{V.~Garc\'{i}a-Sakai}
\affiliation{ISIS Facility, Rutherford Appleton Laboratory, Chilton, Didcot OX11 0QX, United Kingdom}
\author{J.~W. Taylor} 
\affiliation{ISIS Facility, Rutherford Appleton Laboratory, Chilton, Didcot OX11 0QX, United Kingdom}
\author{C.~D.~Frost} 
\affiliation{ISIS Facility, Rutherford Appleton Laboratory, Chilton, Didcot OX11 0QX, United Kingdom}
\author{D.~Prabhakaran} 
\affiliation{Department of Physics, Clarendon Laboratory, University of Oxford, Park Road, Oxford, OX1 3PU, United Kingdom}
\author{C.~MacEwen} 
\affiliation{School of Physics and Astronomy, University of Edinburgh, Edinburgh EH9 3FD, United Kingdom}
\author{A.~Kitada}
\affiliation{Department of Materials Science and Engineering, Kyoto University, Yoshida-honmachi, Sakyo, Kyoto 606-8501, Japan}
\author{A.~J.~Browne} 
\affiliation{School of Chemistry, University of Edinburgh, Edinburgh EH9 3FJ, United Kingdom}
\affiliation{Centre for Science at Extreme Conditions, University of Edinburgh, Edinburgh EH9 3FD, United Kingdom}
\author{M.~Songvilay}
\affiliation{Centre for Science at Extreme Conditions, University of Edinburgh, Edinburgh EH9 3FD, United Kingdom}
\affiliation{School of Physics and Astronomy, University of Edinburgh, Edinburgh EH9 3FD, United Kingdom}
\author{Z. Yamani}
\affiliation{National Research Council, Chalk River, Ontario K0J 1JO, Canada }
\author{W.~J.~L.~Buyers}
\affiliation{National Research Council, Chalk River, Ontario K0J 1JO, Canada }
\affiliation{Canadian Institute of Advanced Research, Toronto, Ontario M5G 1Z8, Canada}
\author{J.~P.~Attfield}
\affiliation{School of Chemistry, University of Edinburgh, Edinburgh EH9 3FJ, United Kingdom}
\affiliation{Centre for Science at Extreme Conditions, University of Edinburgh, Edinburgh EH9 3FD, United Kingdom}
\author{C.~Stock}
\affiliation{Centre for Science at Extreme Conditions, University of Edinburgh, Edinburgh EH9 3FD, United Kingdom}
\affiliation{School of Physics and Astronomy, University of Edinburgh, Edinburgh EH9 3FD, United Kingdom}

\date{\today}

\begin{abstract}

Supplementary information is provided on sample characterization of the MgO powders weakly substituted with magnetic Co$^{2+}$.  The purpose of this characterization is to determine the Co$^{2+}$ concentration and to show the lack of observable Co$^{2+}$ clustering therefore implying a homogenous distribution of magnetic ions on the rocksalt MgO lattice in our samples.  Details on the estimation of the Curie-Weiss temperature used in the main text to compare our results to pure CoO is also given.

\end{abstract}

\maketitle

\renewcommand{\figurename}{Figure \hspace*{-1mm}}
\renewcommand{\topfraction}{0.85}
\renewcommand{\bottomfraction}{0.85}
\renewcommand{\textfraction}{0.1}
\renewcommand{\floatpagefraction}{0.75}

\section{Diffraction}

X-ray diffraction results are shown in Fig. \ref{fig:MvsT} $(a)$ for pure CoO and MgO and our diluted sample of Mg$_{1-x}$Co$_{x}$O.  Rietveld refinement of Mg$_{1-x}$Co$_{x}$O indicates that the solid solution assumes a rocksalt structure ($Fm\bar{3}m$) with a unit cell parameter $a$~=~4.2131(2) \AA. Utilizing the measured values of the end members: CoO (4.2594(4)~\AA) and MgO (4.2118(1)~\AA), the unit cell parameter of 4.2131(2) \AA~corresponds to an $x$~=~0.025(5) according to Vegard's law~\cite{vegard}, confirming that approximately 3\% of the Mg$^{2+}$ sites contain Co$^{2+}$.

\section{Susceptibility}

\textit{Concentration:} For the purposes of consistency, the value of $x$ was also determined \emph{via} DC magnetic susceptibility. As illustrated in Fig.~\ref{fig:MvsT}$(b)$, there is a distinct absence of magnetic ordering between 2 and 300~K, in contrast to CoO. The high temperature portion of the data (T~=[200, 300]~K) exhibits Curie-Weiss behavior with a Curie constant $C$ and Curie-Weiss temperature $\rm{\theta_{CW}}$ of 0.105(8)~emu~$\cdot$~K/mol~Mg$_{0.97}$Co$_{0.03}$O and $-41(6)$~K, respectively. As established in the mean field derivation of the Curie-Weiss law, the Curie constant $C$ must account for all Co$^{2+}$ in Co$_{x}$Mg$_{1-x}$O~and thus its value is directly proportional to $x$. \\   

\indent Recall that the effective paramagnetic moment is defined as 
\begin{equation}
\mu_{eff} = g_{J} \cdot \sqrt{S \cdot (S+1)},
\label{eq:mu_eff} 
\end{equation}    

\noindent where $g_{J}$ is the Land\'{e} g-factor equal to 2 and $S=\frac{3}{2}$.   This would give an expected magnetic moment for Co$^{2+}$ of $\mu_{eff}=3.9 \mu_{B}$.

%Since the $j\rm{_{eff}}$~=~$\frac{1}{2}$ and $j\rm{_{eff}}$~=~$\frac{3}{2}$ manifolds are separated by 36~meV ($\approx$~420~K), the doublet ground state manifold can be effectively thought as thermally isolated in the temperature range probed. Consequently, it is necessary to project all quantities in Eq.~\ref{eq:mu_eff} to the ground state manifold
%\begin{equation}
%\mu_{eff} = g_{J}' \cdot \sqrt{\alpha' S' \cdot (\alpha' S'+1)},
%\end{equation}

%\noindent where primed quantities denote effective values, i.e. $g_{J}'$ and $S'$ denote the effective Land\'{e} g-factor for the ground state doublet ($\frac{13}{3}$) and an effective spin-$\frac{1}{2}$. Inserting the aforementioned values for $g_{J}'$ = $\frac{13}{3}$~\cite{buyers}, $S'$~=~$\frac{1}{2}$ and $\alpha'$~=~$\frac{5}{3}$ where $\alpha'$ denotes the projection factor of $S$ onto a $j\rm{_{eff}}$~=~$\frac{1}{2}$ manifold \emph{via} the projection theorem~\cite{projection}, one obtains

%\begin{equation}
%\mu_{eff} = \frac{13}{3} \cdot \sqrt{\frac{5}{3} \cdot\frac{1}{2} \cdot \left(\frac{5}{3} \cdot \frac{1}{2} +1 \right)} = 5.36~\mu_{B}.
%\label{eq:result}
%\end{equation}

In order to compare the expected value of $\mu_{eff}$ in Eq.~\ref{eq:mu_eff} to the its experimentally determined value, one must first employ the definition of the Curie constant 
\begin{equation}
C = \frac{N\mu_{eff}^{2}}{3k_{B}}.
\label{eq:curie}
\end{equation}

\noindent It should be noted that $\chi$ in Fig.~\ref{fig:MvsT}$(b)$ was normalized by moles and thus $N$ in Eq.~\ref{eq:curie} must be set to $N\rm{_{A}}$. Finally, by solving for $\mu_{eff}$, inserting the value of 0.105(8) emu/K mol Mg$_{0.97}$Co$_{0.03}$O, and utilizing the fact that only 3\% of the Co$^{2+}$ sites are in fact magnetic one obtains an effective paramagnetic moment of 
\begin{equation}
\mu_{eff} = \sqrt{\frac{3k_{B}}{N_{A}} \cdot \frac{1}{0.03} \cdot 0.105(8)} = 5.3 \pm 1.2 ~\mu_{B},  
\label{eq:result2}
\end{equation} 

\noindent in agreement with the predicted value (Eq.~\ref{eq:mu_eff}) and thus confirming that approximately 3\% of the Co$^{2+}$ sites contain Co$^{2+}$.   This is consistent with the starting concentration and also the values estimated from x-ray diffraction discussed above.  The large estimated errorbar in the above equation results from the error in the Co$^{2+}$ concentration and also the mass of the measured sample. 

\renewcommand{\thefigure}{S1}
\begin{figure*}
	\centering
	\includegraphics[width=1.0\linewidth]{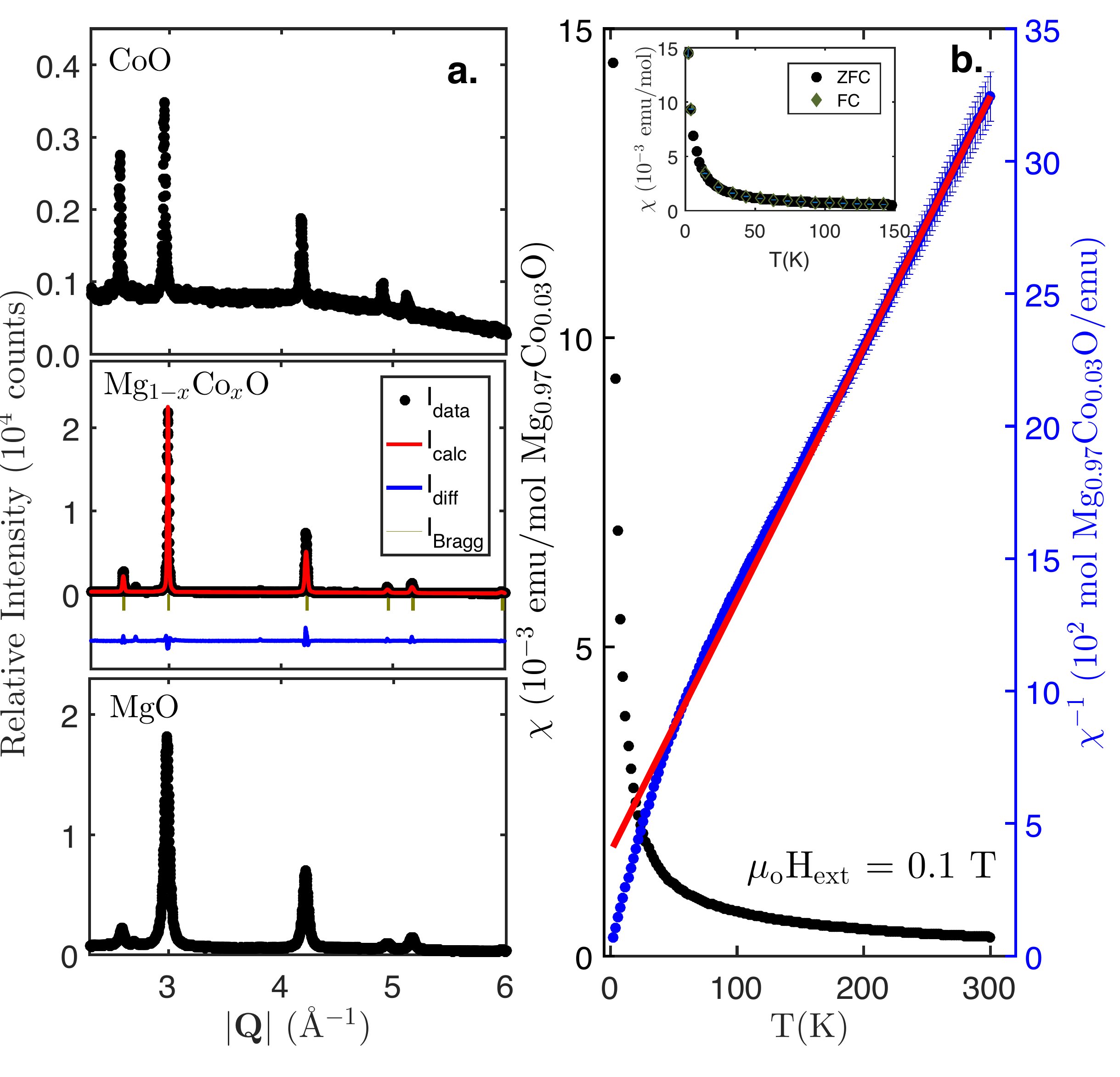}
	\caption{(a) Room temperature diffraction profiles for CoO, Mg$_{1-x}$Co$_{x}$O and MgO collected on a Bruker D2 Phaser x-ray diffractometer utilizing a monochromated Cu K$_{\alpha,1,2}$ source. Rietveld refinement ($\chi^{2}$~=~6.91, $R_{p}$~=~10.12\% and R$_{wp}$~=~13.25\%) of Co$_{x}$Mg$_{1-x}$O indicates that the solid solution assumes a rock-salt structure ($Fm\bar{3}m$) with a unit cell parameter $a$~=~4.2131(2)~\AA, corresponding to an $x$~=~0.025(5) according to Vegard's law~\cite{vegard}. (b) Temperature dependence of molar magnetic susceptibility ($\rm{\mu_{o}H_{ext}}$~=~0.1~T) and its inverse for polycrystalline Co$_{x}$Mg$_{1-x}$O synthesized by sol-gel. ZFC measurements were performed in 2~K steps spaced linearly from 2~K to 300~K, whilst FC measurements were performed in 5~K steps spaced linearly. (inset) A comparison of the temperature dependence of FC and ZFC molar magnetic susceptibility reveals no significant difference indicating an absence of measurable glassy behavior. The high temperature portion of the data (T~=~[200,300]~K) exhibits Curie-Weiss behavior (red-line) with a Curie constant $C$ and Curie-Weiss temperature $\rm{\theta_{CW}}$ of 0.105(8) emu $\cdot$ K/mol~\mgcoo~and $-41(6)$~K, respectively.}
	\label{fig:MvsT}
\end{figure*}

\renewcommand{\thefigure}{S2}
\begin{figure*}
	\centering
	\includegraphics[width=1.0\linewidth]{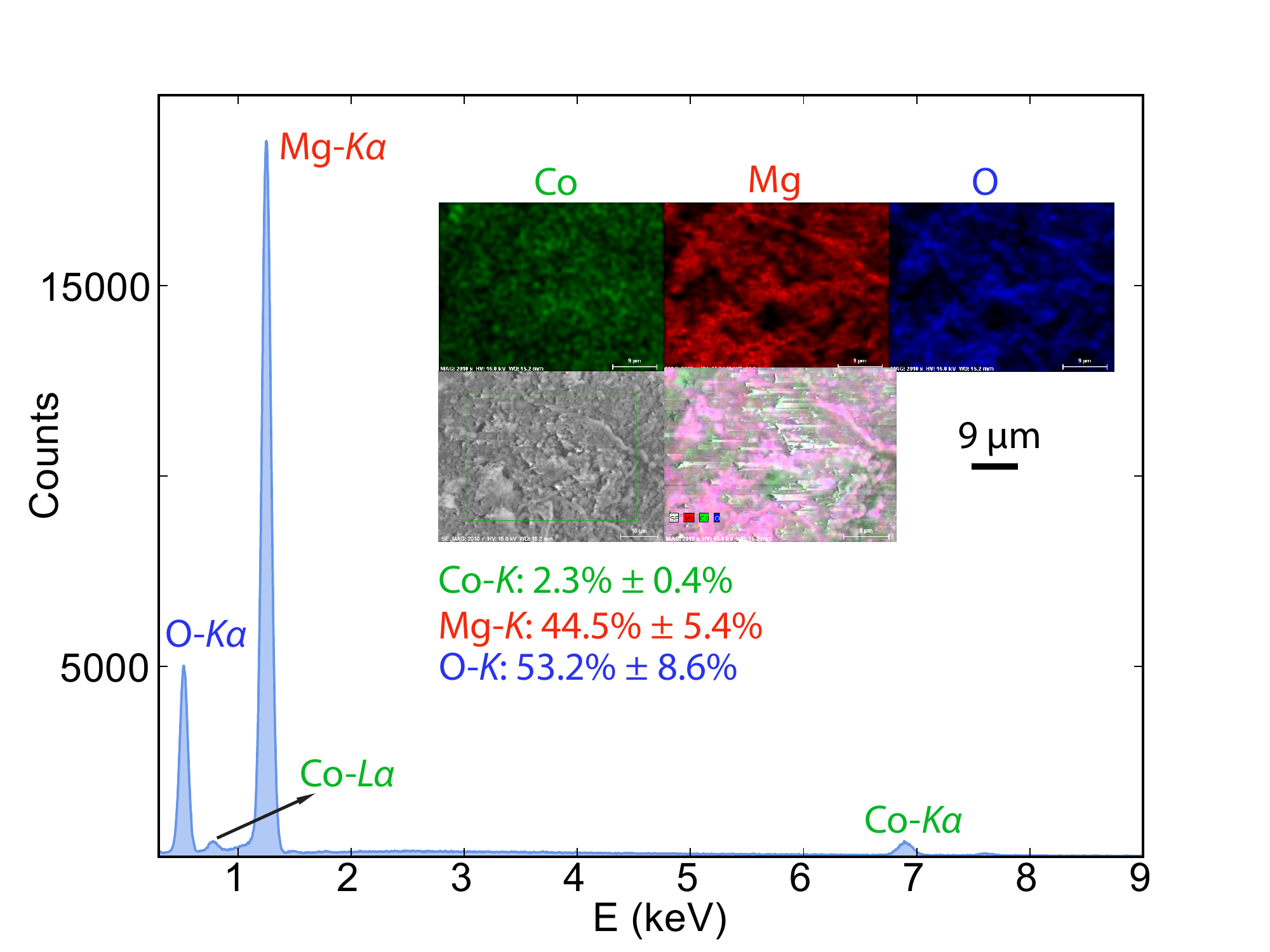}
	\caption{Energy dispersive spectroscopy analysis of the substituted MgO sample used in our neutron experiments.  The resulting elemental analysis is shown and consistent with x-ray and susceptibility measurements.}
	\label{fig:EDX}
\end{figure*}

\textit{Homogeneous concentration of Co$^{2+}$:}   A confirmation of the prevalence of pairs over larger clusters of Co$^{2+}$, and hence a homogeneous distribution of Co$^{2+}$ in the sample, is given the value of the experimentally determined Curie-Weiss constant from DC susceptibility shown in Fig. \ref{fig:MvsT} $(b)$. It is rationalized based on probability that Co$^{2+}$ in Mg$_{0.97}$Co$_{0.03}$O are predominately isolated (i.e. single ions), while a smaller but significant amount of Co$^{2+}$ exhibit pairwise interactions. These pairs are coupled by $J_{i}$ where the index $i$ denotes a particular coordination shell.  A confirmation of the presence of pairs can be obtained from Curie-Weiss temperature which in the mean-field limit has the form,

	\begin{equation}
	\theta_{CW} = -\frac{2S(S+1)\sum\limits_{i}z_{i}J_{i}}{3},
	\label{eq:CW}
	\end{equation} 
	
\noindent where $z_{i}$ denotes the number of atoms/ions coupled to the central atom/ion with magnetic constant $J_{i}$~\cite{MFtheory}. The concurrent presence of AFM and FM behavior in each coordination shell, with the exception of the second, implies that $\rm{\theta_{CW}}$ would be dominated by the second coordination shell. Before proceeding, it is worth noting that Kanamori~\cite{1957} utilizing second order perturbation theory deduced that strong spin-orbit coupling in CoO substantially reduces the Curie-Weiss temperature estimated by the mean field approximation in Eq.~\ref{eq:CW}. Utilizing a value of approximately 0.9 eV for the energy splitting between the $^{4}F$ and $^{4}P$ free ion states, Kanamori derived that his mean field estimate of $\theta\rm{_{CW}} \sim$ $-530$ K was overestimated by a factor of 1.9 if spin-orbit coupling was neglected. Thus, by inserting the values of $S$~=~$\frac{3}{2}$, $J_{2}$~=~35.9(6)~K (3.09(5)~meV), setting the value of $z$ as 1 since individual Co$^{2+}$ pairwise interactions are of interest and dividing by a correction factor of 1.9 deduced by Kanaomori~\cite{1957} to account for spin-orbit coupling yields a value of 
	
	\begin{equation}
	\widetilde{\theta}_{CW} = -\frac{2 \cdot \frac{3}{2} (\frac{3}{2}+1)\cdot(1)(35.9(6))}{3\cdot1.9} = -44.9(7)~\rm{K},
	\label{eq:CW2}
	\end{equation}
	
\noindent in excellent agreement with the experimentally determined value of $-41(6)$~K for our dilute samples shown in Fig. \ref{fig:MvsT} $(b)$. The success in accounting for the Curie-Weiss temperature implies that the cooperative magnetism of Mg$_{0.97}$Co$_{0.03}$O can be approximated as being attributed to Co$^{2+}$ pairwise interactions exclusively. As a concluding comment, it is worthwhile to note that the magnetic susceptibility exhibits no difference between ZFC/FC, implying an absence of glassy behavior down to 2~K. This absence of glassy behavior supports our claim that larger molecular clusters are absent since these would be expected to yield glassy behavior and thus a ZFC/FC split.  

\section{Energy dispersive x-ray analysis:}  Energy dispersive x-ray analysis results are displayed in Fig. \ref{fig:EDX}.  As illustrated in the figures, the concentration of magnetic Co$^{2+}$ was measured to be 2.3 \% $\pm$ 0.4 \%, consistent with susceptibility and diffraction results discussed above.  The inset figures to Fig. \ref{fig:EDX} also show the distribution to be uniform with no obvious clustering consistent with our susceptibility analysis above.

\renewcommand{\thefigure}{S3}
\begin{figure*}
	\centering
	\includegraphics[width=1.0\linewidth]{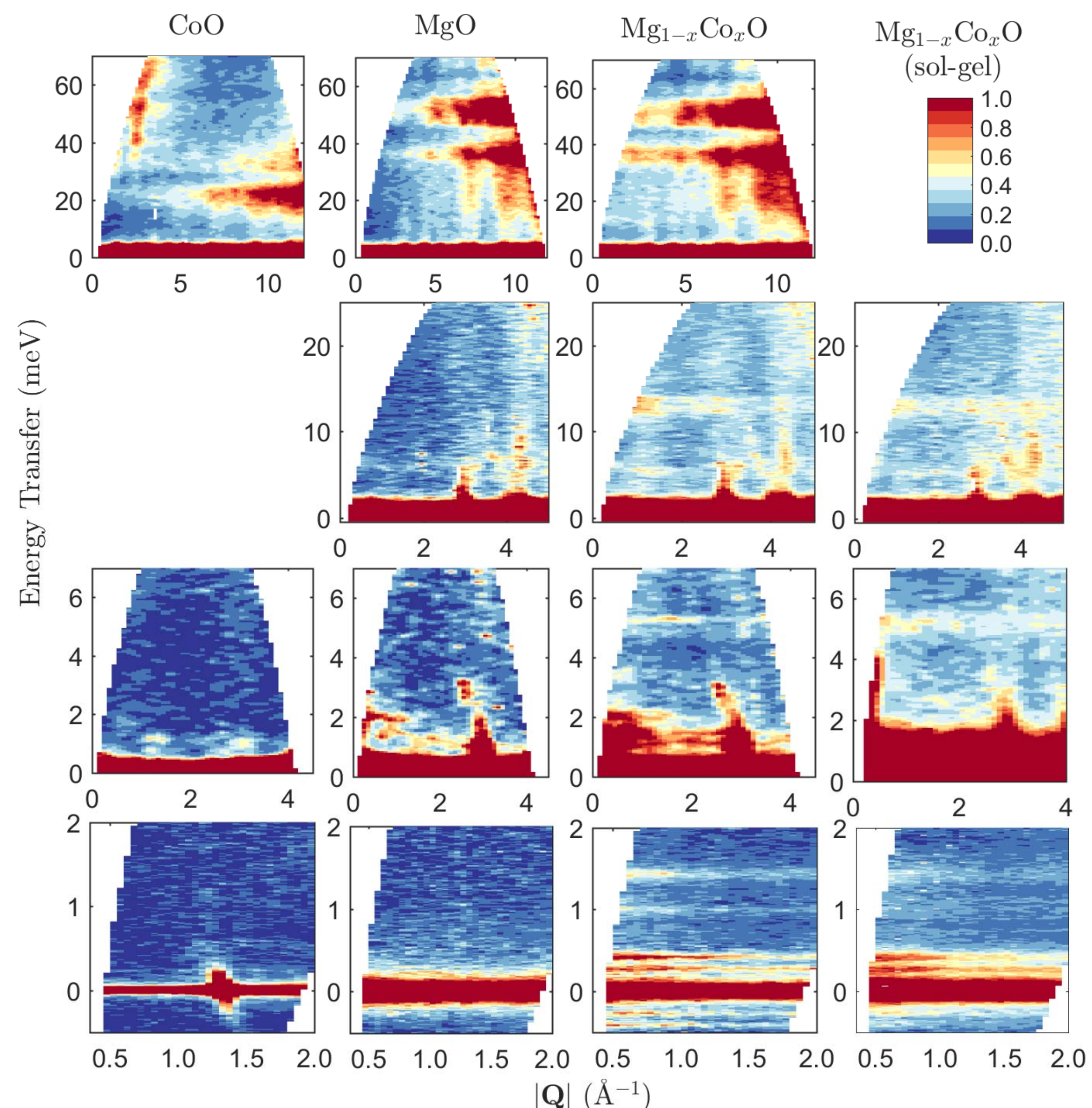}
	\caption{A comparison of raw (non-background subtracted) $I(|\mathbf{Q}|,E)$ maps at 5~K for CoO, MgO, Mg$_{0.97}$Co$_{0.03}$O synthesized by standard solid state methods and Mg$_{0.97}$Co$_{0.03}$O synthesized by sol-gel. Each column corresponds to a particular compound as labelled from above. Rows 1 to 3 correspond to incident energies of 85~meV ($f$~=~300~Hz), 30~meV ($f$~=~350~Hz), 10~meV ($f$~=~250~Hz) measured on MARI, respectively, whilst row 4 corresponds to a final energy of 1.84~meV measured on IRIS. Three exceptions include: CoO in row 1 corresponding to an incident energy of 100~meV ($f$~=~350~Hz), Mg$_{0.97}$Co$_{0.03}$O (sol-gel) in rows 2 and 3 correspond to incident energies of 29.5~meV ($f$~=~350~Hz) and 14.5~meV ($f$~=~250~Hz), respectively. All $I(|\mathbf{Q}|,E)$ maps have been renormalized to a common scale.}
	\label{fig:figureS1}
\end{figure*}

\section{Spectroscopic comparison of pure and substituted samples}

We compare the spectroscopic response of pure CoO, MgO, and our two dilute samples prepared with solid state and solution techniques in Fig. \ref{fig:figureS1}.  The two different sample preparation techniques provide consistent spectroscopic results for all energy transfers measured.  The results are different from those obtained in pure CoO and MgO indicative of a lack of local large clusters of Co$^{2+}$ magnetic moments.  Based on this comparison, we conclude that the results presented in this paper are independent of sample preparation techniques and also that our samples do not contain local regions of magnetic clustering.

\section{Tabulated experimental energy positions:}

In this section we list the measured experimental energy positions and the calculated coordination shell ($m$) along with the calculated $J$ including the sign indicative of ferromagnetic or antiferromagnetic coupling.  The results are used below for the analysis of the Curie-Weiss constant.

\begin{table*}
	\caption{Tabular summary of the calculated parameters related energy peak position to exchange constant}  	
	\begin{ruledtabular}
		\begin{tabular}{ccccc}
			%\Xhline{3\arrayrulewidth} 
			Experimental $\Delta E_{o}$ (meV)~~& Calculated $|\mathbf{R}|$ (\AA)~~&Relative Coordination Shell $m$& Type of Coupling & Calculated $J$ (meV) \\ [1ex] % inserts table
%heading
\hline % inserts single horizontal line  
13.1(2) 	& 4.2(3) &2  & AF  &  3.09(5)  \\ \hline 
5.256(4) & 4.1(5) &1  & AF  & 1.000(8)    \\ \hline 
4.857(3) & 4.4(3) &1  & F  &  -0.918(6)  \\ \hline 
1.434(3) & 5.5(5) &3  & AF  & 0.258(1)   \\ \hline 
0.998(4) & 5.4(6) &3  & F  &   -0.182(1) \\ \hline 
0.420(2) & 6.3(7) &4  & AF  & 0.0759(4)   \\ \hline 
0.279(2) & 5.8(5) &4 & F &  -0.0504(4) \\   
		\end{tabular}
	\end{ruledtabular}
	\label{tab:1}
\end{table*}

\begin{table*}
	\caption{Select summary of crystallographic parameters of \mgcoo. The number of neighbors in a relative coordination shell $m$ was determined assuming a collinear type-II antiferromagnetic structure possessed by CoO~\cite{jauch,DFT}.} 	
\begin{ruledtabular}
	\begin{tabular}{ccccc}
		%\Xhline{3\arrayrulewidth} 
			Relative Coordination Shell $m$ & Distance at 298 K $|\mathbf{R}_{m}|$ (\AA) & Number of Neighbors in $m$ & AF Coupling & F Coupling   \\ [1ex] % inserts table
		%heading
		\hline % inserts single horizontal line  
			1 	&   2.983(1) & 12 & 6 & 6\\ \hline 
2 &  4.218(2) & 6 & 6 & 0\\ \hline 
3 &  5.166(2) & 24 & 12 & 12\\ \hline 
4 &  5.966(2) & 12 & 0 & 12\\ 
	\end{tabular}
\end{ruledtabular}
	\label{tab:0}
\end{table*}

\section{Estimation of the Curie-Weiss temperature for CoO \emph{via} a mean field approach:}

The interpretation of the magnetic excitations illustrated in Fig.~2$(a)$ in the main text as distinct pairwise interactions due to the assumption of physical homogeneity in \mgcoo~corresponds to a direct analog of the physical model underlying the mean field estimate of the Curie-Weiss temperature given by Eq.~\ref{eq:CW}. Before $\theta\rm{_{CW}}$ can be estimated with Eq.~\ref{eq:CW}, it is important to note that while the value of $S =\frac{3}{2}$ is well-established as determined by Hund's rules~\cite{cowley,sakurai,buyers,cov2o6}, the value of $z_{i}$ possesses an inherent ambiguity in \mgcoo. Since \mgcoo~itself does not magnetically order down to 2~K~\cite{cowley} as illustrated in Fig.~\ref{fig:MvsT}, the number of Co$^{2+}$ exhibiting each type of coupling $J_{i}$ as measured by inelastic neutron spectroscopy cannot be determined directly by referring to its magnetic structure. 

We have assumed that the interactions in \mgcoo~are approximately analogous to those in CoO. Utilizing the assumption of the equivalence between \mgcoo~and CoO, the values of $z_{i}$ can be determined by investigating the number of couplings between different coordination shells in the collinear type-II antiferromagnetic structure proposed for CoO~\cite{DFT,jauch}. The determination of the value of $z_{i}$ for coordination shells $m$~=~1~\dots~4 was accomplished by first applying a transformation of the cubic $Fm\bar{3}m$ space group into its rhombohedral maximal subgroup $R\bar{3}m$ of the hexagonal crystal family. This transformation results in the stacking of the $\langle 111 \rangle$ planes along $c$ in the $R\bar{3}m$ representation. Since Co$^{2+}$ in 0, $\pm$2, \dots and $\pm$1, $\pm$3, \dots layers relative the $\langle 111 \rangle$ layer of reference are coupled F and AF, respectively in a collinear type-II antiferromagnet, then the $c$ coordinate of each Co$^{2+}$ in a particular $m$ coordination shell determined the type of coupling each Co$^{2+}$ was predicted to exhibit. It should be noted that this analysis suggests only ferromagnetic correlations should exist in $m$~=~4 coordination shell, in contradiction with the AF assignment to the magnetic excitation at $\Delta E\rm{_{o}}$~=~0.0759(2)~meV, which may indicate the presence of magnetic frustration. The number of Co$^{2+}$ in each coordination shell exhibiting F or AF coupling is summarized in Tab.~\ref{tab:0}. 

By inserting the values of $S$ as determined from Hund's rules, $J_{i}$ as determined from the measured energy transfers and $z_{i}$ as determined from the aforementioned rhombohedral transformation analysis, into the mean field estimate of the Curie-Weiss temperature with Kanamori's second perturbation order corrections, one obtains 

\begin{align} \label{eq:first_estimate}
\widetilde{\theta}_{CW} &= \left(\frac{-5}{2 \cdot 1.9}\right)  \left\{-0.918(6) \cdot 6  + 1.000(8) \cdot 6 + 3.09(5) \cdot 6 \right. \nonumber \\
&\mathrel{\phantom{=}} \left.\kern-\nulldelimiterspace -0.182(1) \cdot 12 + 0.258(1) \cdot 12 -0.0504(4) \cdot 12  \right\} \nonumber \\
&\mathrel{\phantom{=}} \left.\kern-\nulldelimiterspace  \right. = -295(5) \, \rm{K}, 
\end{align}       

\noindent a value very similar to the experimental value of $\theta\rm{_{CW}}$~=~$-330$~K~\cite{weiss} and in particular, T$\rm{_{N}}$~=~291~K~\cite{jauch}.

%\bibliography{references_Co}
